%% file: reactor_AR_main.tex
\documentclass[12pt]{article}
\usepackage{amsmath,float,amssymb,amsthm,amsxtra,overpic,bbm,bm,epsfig,color}

\def\thefootnote{\fnsymbol{footnote}}

\def\tabnotefont{\fontsize{9}{10}\selectfont}%
\newenvironment{tabnote}{\par\tabnotefont}{\par}

\usepackage{slashed,stmaryrd}
\usepackage{color}

\usepackage[numbers]{natbib}
\usepackage[colorlinks=true,citecolor=blue,linkcolor=blue]{hyperref}% PDF links
\usepackage{amsmath,amssymb}
\usepackage{graphicx,color}
\usepackage{sidecap}
\usepackage{footnote}
\makesavenoteenv{tabular}
\usepackage{xspace}

\newcommand{\Linine}     {$^{9}$Li\xspace}
\newcommand{\Btwelve}     {$^{12}$B\xspace}
\newcommand{\Heeight}      {$^{8}$He\xspace}
\newcommand{\zeronubb}    {$0\nu\beta\beta$\xspace}

\newcommand{\mbb}         {$m_{\beta\beta}$\xspace}

\newcommand{\nuebar}      {$\overline\nu_e$\xspace}

\begin{document}
\vspace{0.2cm}

\begin{center}
{\Large\bf Reactor Neutrino Experiments: Present and Future}
\end{center}

\vspace{0.2cm}

\begin{center}
{\bf L.~J.~Wen $^{a}$}, \footnote{E-mail: wenlj@ihep.ac.cn}
\quad
{\bf J.~Cao $^{a}$}, \footnote{E-mail: caoj@ihep.ac.cn}
\quad
{\bf Y.~F.~Wang $^{a}$}, \footnote{E-mail: yfwang@ihep.ac.cn}
\\
{\small $^a$Institute of High Energy Physics, Chinese Academy of
Sciences, Beijing 100049, China}
\end{center}

\vspace{1.5cm}

\begin{abstract}
Reactor neutrinos have been an important tool for both discovery and precision measurement in the history of neutrino studies. Since the first generation of reactor neutrino experiments in the 1950s, the detector technology has been greatly advanced. New ideas, new knowledge, and modern software also enhanced the power of the experiments. The current reactor neutrino experiments, Daya Bay, Double Chooz, and RENO have led neutrino physics into the precision era. In this article, we will review these developments and accumulations, address the key issues in designing a state-of-art reactor neutrino experiment, and explain how the challenging requirements of determining the neutrino mass hierarchy with the next generation experiment JUNO could be realized in the near future.
\end{abstract}

%Keywords, etc.
\begin{flushleft}
\hspace{0.9cm} Key words: reactor neutrino, neutrino oscillations, mass hierarchy, lepton flavor
\end{flushleft}

\def\thefootnote{\arabic{footnote}}
\setcounter{footnote}{0}

\newpage

%\tableofcontents

\input{section_1}

\input{section_2}
\input{section_3}
\input{section_4}
\input{section_5}
\input{section_6}

\vspace{12pt}
\section*{ACKNOWLEDGMENTS}
We thank Dr. Jie Zhao for producing Fig.~1, Dr. Fengpeng An for producing Fig.2 and Xiaohui Qian for producing Fig.~6. The writing of this review was supported by the Strategic Priority Research Program of the Chinese Academy of Sciences, Grant No. XDA10010900; the CAS Center for Excellence in Particle Physics (CCEPP) (for all authors).

\end{document}

%% file: section_1.tex
%!TEX root = reactor_AR_main.tex
%%%%%%%%% Section: Intro %%%%%%%%%

\section{INTRODUCTION}

\label{sec:intro}
\par
Reactor neutrino experiments have played a critical role in the 60-year-long history of neutrinos. After the discovery of the neutrino by Reines and Cowan~\cite{Cowan56} in 1956, early searches for oscillation with reactor neutrinos in late 1970s and 1980s turned out to be controversial, but led to much better understanding of the reactor neutrino flux. Palo Verde~\cite{Paloverde} and CHOOZ~\cite{Chooz} introduced modern particle physics technology into reactor neutrino experiments such as detector Monte Carlo and complex veto design in the 1990s. Detector systematics were well understood although no oscillation at $\sim\,$1 km was found. KamLAND~\cite{Kamland03} found reactor neutrino oscillation at a 180 km baseline and tackled the degeneracy in solar neutrino oscillation parameters in 2002. Backgrounds studies were well advanced. Based on these experiences, Daya Bay~\cite{Dayabay}, Double Chooz~\cite{DChooz}, and RENO~\cite{Reno} were designed with unprecedented precision in 2000s and finally measured the smallest mixing angle $\theta_{13}$ by observing the oscillation at $\sim\,$1 km, which led the neutrino physics into the precision era. The next generation experiment JUNO, expected to operate in 2020, will further advance the capability of reactor neutrino experiments to determine the neutrino mass ordering and precisely measure several oscillation parameters~\cite{An:2015jdp}.

\par
Reactor neutrinos are electron antineutrinos\footnote{Since there is no exception, reactor neutrino mentioned in this article always means electron antineutrino.} that are emitted from subsequent $\beta$-decays of instable fission fragments. More than 80\% of commercial reactors are light water reactors, which will be used as an example in this article. In these reactors, fission of four fuel isotopes, $^{235}\rm U$, $^{238}\rm U$, $^{239}\rm Pu$, and $^{241}\rm Pu$, makes up more than 99.7\% of the thermal power and reactor antineutrinos. About 6 neutrinos in 0-10 MeV range are released from each fission of these isotopes, together with $\sim\,$200 MeV energy.

\par
The inverse beta decay (IBD) reaction, $\overline\nu_e + p \rightarrow n+e^+$, having the largest cross section in a few MeV range and incomparable power to reject backgrounds, is the classical channel to detect reactor neutrinos with liquid scintillator (LS) rich in hydrogen. The positron carries almost all the energy of the antineutrino and forms a {\it prompt} signal. After thermalization, the neutron is captured by a hydrogen or other nuclei (e.g. Gd, Cd, Li) and forms a {\it delayed} signal, tens or hundreds microseconds later than the {\it prompt} signal. Roughly the event rate is $\sim 1/({\rm ton}\cdot {\rm GW_{th}} \cdot {\rm day})$ at 1 km distance from the reactor, where ton is the unit of the target mass of the liquid scintillator and GW$_{\rm th}$ is the unit of the thermal power of the reactor. Elastic scattering of reactor neutrino on electron was used to study the magnetic moment of neutrino with Germanium detectors~\cite{Wong:2006nx, Beda:2013mta} or gas TPC detector~\cite{Daraktchieva:2005kn}, but deviates from the focus of this article.

\par
In Sec.~\ref{sec:history} we will briefly describe the major reactor neutrino experiments. Key issues in designing a state-of-art reactor neutrino experiment will be described in Sec.~\ref{sec:expTech}, followed by the design of JUNO and new technical advancements in Sec.~\ref{sec:futureExp}, as well as a short summary in Sec.~\ref{sec:conclusion}.

%% file: section_2.tex
%!TEX root = reactor_AR_main.tex
\section{REACTOR NEUTRINO EXPERIMENTS}
\label{sec:history}

\subsection{Early experiments}

\par
The first proposal to detect the free neutrino was to use nuclear bombs. Then it was superceded by the approach of using a fission reactor~\cite{Reines:1953kf} with the technology of liquid scintillator in the 1950s~\cite{Cowan:1953mw}. The Hanford experiment~\cite{Reines53} made the first attempt to search for the neutrino. Shielded by paraffine and lead, its 300 liter liquid scintillator was viewed by 90 2-inch PMTs. No neutrino signal was found due to high backgrounds. The detectors were moved back to Los Alamos and put underground, which confirmed the backgrounds were from cosmic rays. The lesson was clear: ``it is easy to shield out the noise men make, but impossible to shut out the cosmos". Therefore, the detector has to be put underground.

\par
Besides the passive shielding of the cosmic rays, the first anti-coincidence detector was developed in the later Savannah River experiment to distinguish backgrounds. The antineutrino signature was a coincidence between the prompt $e^+$ annihilation signal and the microseconds delayed neutron capture on cadmium. The detector was deployed close ($\sim\,$11 m) to the Savannah River reactor and 12 m underground in a massive building. Finally the existence of neutrinos were convinced~\cite{Cowan56}.

\par
Immediately after the observation of neutrinos, Pontecorvo~\cite{Pontecorvo57,Pontecorvo58}, Maki, Nakagawa and Sakata~\cite{MNS62} suggested that the neutrino may oscillate from one flavor to another as it travels, which stimulated searches for neutrino oscillations. The first experiment on neutrino oscillations was performed in 1979 near Savannah River~\cite{Reines:1980pc}. It was the first experiment that use heavy water to detect neutrinos~\cite{Pasierb:1979fb}, by neutral-current interactions (NC: $\overline\nu_e+d\rightarrow n+p+\overline\nu_e$) and charged-current interactions (CC: $\overline\nu_e+d\rightarrow n+n+e^+$). The neutrino oscillation signal, expressed as the double ratio of the measured to theoretical rate of the CC with respect to NC process, was measured to be 0.40$\pm$0.22, a first indication of neutrino instability. It was interpreted as neutrino oscillation and deduced an allowed region of $\Delta m^2$ and $\sin^22\theta$ assuming two base mass states to be involved.

\par
A long debate followed about the indication of neutrino oscillation. The ILL (Institut Laue-Langevin) experiment~\cite{Boehm:1980pm,ILL} in France used 377 liter LS placed at 8.75 m from the ILL reactor (57 MW, 93\% $^{235}$U). For the first time the reactor neutrino spectrum was calculated~\cite{Davis:1979gg,Vogel:1980bk}. No evidence of oscillation was found, as the ratio of measured neutrino events to the theoretical prediction was 0.89$\pm$0.04(stat.)$\pm$0.14(syst.)~\cite{Davis:1979gg}. The upper limits for the $\Delta m^2$ and $\sin^22\theta$ obtained from ILL excluded the allowed region from the above Savannah River experiment.

\par
The Bugey experiment~\cite{Cavaignac:1984sp} in France made a new claim. The positron energy spectra were measured at 13.6 m and 18.3 m from the Bugey reactor with a detector similar to the ILL experiment. Their ratio was regarded as less depending on the knowledge of the initial spectrum or other detector systematics. It found indication of neutrino disappearance at the 3$\sigma$ level by using the ratio of the positron spectra at two baselines.

The detector from the ILL experiment was upgraded and transferred to a more powerful (2.8 GW$_\text{th}$) commercial reactor at G$\ddot{o}$sgen~\cite{Gosgen}, Switzerland. The \nuebar signals were measured at distances of 37.9 m, 45.9 m, and 64.7 m from the reactor core. Very good agreement between data and expectation, both in rate and spectrum, was obtained. Almost the entire parameter region suggested by the Bugey experiment was excluded. Later the Bugey-3 experiment~\cite{Declais:1994su} deployed new detector modules at 15 m, 40 m and 95 m from the Bugey reactor to search for neutrino oscillations. Again, no oscillations were observed, and the exclusion region in the oscillation parameter space was largely extended.

\subsection{CHOOZ and Palo Verde}
\par
Parallel to the Super-Kamiokande experiment which finally discovered the neutrino oscillation with atmospheric neutrinos~\cite{SuperK98}, Palo Verde~\cite{Paloverde,Boehm:2000vp,Boehm:2001ik} and CHOOZ~\cite{Chooz,Apollonio:2002gd} experiments were built to test the hypothesis that neutrinos oscillate with the parameters indicated by the atmospheric neutrino measurements.

\par
The Palo Verde detector was built at distances of 750, 890 and 890 m from the three reactors of the Palo Verde Nuclear Generating Station. The shallow overburden is only 12 m ($\approx$ 32 m.w.e), thus it just reduces the muon flux by a factor of five, resulting in substantial muon-induced neutrons. To suppress the neutron background, a unique segmented detector was designed, with a target consisting of 11$\times$6 cells filled with Gadolinium (Gd) loaded liquid scintillator. The IBD signature was by a {\it triple} coincidence of one high energy trigger, due to the positron ionization or neutron capture cascade core, and at least two additional low energy triggers, resulting from positron annihilation $\gamma$s or neutron capture $\gamma$s. Despite this, the neutron backgrounds are still the overwhelming source of uncertainties and a $swap$ method was specially designed to suppress backgrounds~\cite{Wang:2000kt}.

The CHOOZ experiment opened up the uniform detector at {\it deep} underground. The detector was built in an underground cavity with an overburden of 300 m.w.e, near two reactors with a total power of 8.5 GW$_{\rm th}$. It is a homogeneous detector and the antineutrino signals are {\it double} coincidences. Comparing to the segmented design of Palo Verde, the homogeneous design of CHOOZ gave full-absorption peak of 8 MeV $\gamma$s from n-capture on Gd, leading to high efficiency and smaller uncertainty on neutron detection. The fast neutron backgrounds were much less than that in Palo Verde due to 10 times larger overburden. Since the PMTs of CHOOZ directly contacted with the scintillator, the radiation from PMT glass, particularly $^{40}$K, resulted in large low energy backgrounds. Furthermore, the varying energy threshold caused by the fast degradation of its gadolinium-loaded scintillator resulted in a large efficiency uncertainty.

\par
The negative results of Palo Verde and CHOOZ concluded that the atmospheric neutrino oscillations do not involve $\nu_e$, and set an upper limit for $\sin^22\theta$ under the 2-flavor oscillation scenario~\cite{Boehm:2000vp}. In the case of $\Delta m^2=0.0024$ eV$^2$, the limit was $\sin^22\theta_{13}$\textless0.12 at 90\% C.L.

\par
Compared to the first generation experiments, Palo Verde and CHOOZ introduced modern particle physics technology into reactor experiments, such as precise detector calibration, active veto detector, and Monte Carlo techniques (e.g, GEANT3 and FLUKA), which significantly improved the understanding of detector systematics and backgrounds.

\subsection{KamLAND}
\par
The ``Solar Neutrino Problem'' refers to the persistence of a large deficit of the solar $\nu_e$ flux relative to the Standard Solar Model (SSM)~\cite{Bahcall}, found by a series of solar neutrino experiments~\cite{Homestake,GALLEX,SAGE,Kamiokande,Super-Kamiokande} in the late 1960s. It was tempting to invoke explanations based on the neutrino oscillation hypothesis, and these experiments allowed several possible solutions in the oscillation parameter space of $\sin^22\theta$ and $\Delta m^2$. The solutions are usually referred to as large mixing angle solution (LMA), small mixing angle solution (SMA), low $\Delta m^2$ solution (LOW) and vacuum (VAC) oscillation. In 2001, the SNO solar neutrino experiment gave the ``smoking gun" evidence of the neutrino oscillation explanation to the Solar Neutrino Problem. However, the above four solutions are flux (SSM) dependent thus flux independent evidence is needed.

\par
A multi-purpose detector, KamLAND, is placed at the site of the former Kamiokande experiment with a vertical overburden of 2,700 m.w.e, with the primary goal to search for reactor $\overline\nu_e$ oscillations. It is surrounded by 55 Japanese nuclear reactor cores. The $\overline\nu_e$ flux weighted average baseline is $\sim\,$180 km. KamLAND consists of 1 kton of ultrapure liquid scintillator in a 13-m-diameter transparent spherical balloon. The balloon is supported by ropes in a buffer of dodecane and isoparaffin oils between the balloon and a 18-m-diameter stainless steel (SS) sphere. A non-rigid balloon led to a not well-defined target mass, and 2.1\% uncertainty in the total LS mass was determined by the measurement with flow meters during detector filling. Due to high backgrounds from the balloon, a fiducial volume cut was applied, resulting in a large error (4.1\%) which relied on the vertex reconstruction. Later a 4$\pi$ calibration system~\cite{Busenitz:2009ac} significantly reduced the systematic errors. The vertex bias was reduced from 5 cm to 3 cm, leading to an improvement of fiducial volume uncertainty from 4.7\% to 1.8\%~\cite{Kamland08}.

\par
The KamLAND results~\cite{Kamland03,Araki:2004mb,Kamland08} are highly consistent with the solar neutrino experiments, and have confirmed the LMA solution to be the solution of the Solar Neutrino Problem. The combination of SNO and KamLAND gave the most precise measurements of $\tan^2\theta_{12}=0.47^{+0.06}_{-0.05}$ and $\Delta m^2_{21}=7.59^{+0.21}_{-0.21}\times 10^{-5}$eV$^2$.

\subsection{Daya Bay, Double CHOOZ and RENO}

\par
As neutrino oscillation was well established around 2002, the mixing parameter $\theta_{23}\sim45^\circ$ was determined by the atmospheric~\cite{SuperK98} and long-baseline accelerator~\cite{K2K-2003} neutrino experiments, and $\theta_{12}\sim33^\circ$ was determined by the solar neutrino experiments and KamLAND. The third mixing angle $\theta_{13}$ was still unknown. An upper limit has been determined by CHOOZ and Palo Verde. Leptonic CP violation is an effect under 3-flavor framework and can only be tested if $\theta_{13}$ is non-zero. If $\theta_{13}>0.01$, the chance exists to determine the neutrino mass ordering and measure the CP phase in next generation experiments.

\par
%A relative measurement with two detectors, one close to the reactor and another at the oscillation maximum, was proposed in 1999 to cancel the uncertainties from reactors and correlated uncertainties from detectors~\cite{Mikaelyan:1999pm}. The proposal was not taken seriously until strong motivations came forth to measure $\theta_{13}$ around 2003. Eight proposals came out, including Angra~\cite{Anjos:2005pg}, Braidwood \cite{Bolton:2005yd}, Daya Bay~\cite{Guo:2007ug}, Diablo Canyon~\cite{Anderson:2004pk}, Double Chooz~\cite{Ardellier:2006mn}, Kransnoyarsk~\cite{Martemyanov:2002td}, KASKA~\cite{Aoki:2006bk} and RENO~\cite{Ahn:2010vy}. Finally three of them, Daya Bay, Double Chooz, and RENO were built. The advancements of suppressing the correlated-uncertainties via the relative measurement are shown in Table~\ref{tab:ExpUncer}, with the projected uncertainties after Near-Far cancellation taken from Ref.~\cite{Guo:2007ug,Ardellier:2006mn,Ahn:2010vy}. The typical systematic uncertainty of the reactor experiments before 2000 was 3$\sim\,$6\%. Daya Bay, Double Chooz and RENO all aimed at sub-percent level systematics. Especially, Daya Bay was designed at a $\sim\,$0.4\% uncertainty in order to achieve a sensitivity of 0.01 (90\% C.L.) in $\sin^22\theta_{13}$. // this is the original version in the submitted file

A relative measurement with two detectors, one close to the reactor and another at
the oscillation maximum, was proposed in 1999 to cancel the uncertainties from reactors
and correlated uncertainties from detectors (42). This possibility was not taken seriously
until strong proposals were put forth to measure $\theta_{13}$ around 2003. Eight experiments were proposed: Angra~\cite{Anjos:2005pg}, Braidwood \cite{Bolton:2005yd}, Daya Bay~\cite{Guo:2007ug}, Diablo Canyon~\cite{Anderson:2004pk}, Double Chooz~\cite{Ardellier:2006mn}, Kransnoyarsk~\cite{Martemyanov:2002td}, KASKA~\cite{Aoki:2006bk}, and RENO~\cite{Ahn:2010vy}. Finally three of them--Daya Bay, Double Chooz, and RENO--were built.
The near-far cancellation has been crucial for these three experiments that all aimed for sub-percent-level systematics, in particular, Daya Bay was designed at an $\sim$0.4\% uncertainty in order to achieve a sensitivity of 0.01 (90\% CL) in $\sin^22\theta_{13}$. There are three main sources of systematic uncertainties: reactor, background, and detector. More discussions about the uncertainties are in Section 3.3 and Section 4. Table 1 shows the advances in suppressing correlated uncertainties via relative measurements (the projected uncertainties after near-far cancellation are taken from References 45,47,and 50).

\begin{table}[!htb]
\begin{center}
\caption{Detector-related and reactor-related uncertainties from CHOOZ, Palo Verde, KamLAND, and the projected uncertainties by the Near-Far relative measurement.}\label{tab:ExpUncer}
\begin{tabular}{c|c|c|c|c}
\hline
Uncertainties & CHOOZ  & PALO VERDE & KamLAND & Near-Far   \\\hline\hline
Reaction cross section & 1.9\%  &    &   &  0  \\
\cline{1-2}\cline{5-5}
Energy released per fission & 0.6\% & 2.1\%$^{\rm a}$  & 3.4\%$^{\rm b}$   &  0 \\
\cline{1-2}\cline{5-5}
Reactor power & 0.7\%   &    &   &  $\sim\,$0.1\% \\\hline
Number of protons & 0.8\%  &  0.8\%  &    &  \textless0.3\%  \\
\cline{1-3}\cline{5-5}
Detection efficiency & 1.5\%  &  2.1\% &  2.4\%$^{\rm c}$   &  0.2\%-0.6\%  \\\hline\hline
Combined & 2.7\%  &  3.1\% &  4.2\%  &  \textless(0.4\%-0.6\%) \\\hline
\end{tabular}
\begin{tabnote}
  $^{\rm a}$ The total reactor-related uncertainty taken from Ref.~\cite{Boehm:2001ik}.
  $^{\rm b}$ It combined the reactor power, fuel composition, \nuebar-spectra and long-lived nuclei uncertainties taken from Ref.~\cite{Kamland08}.
  $^{\rm c}$ It combined the fiducial volume, energy threshold, efficiency and IBD cross section uncertainties taken from Ref.~\cite{Kamland08}.
\end{tabnote}
\end{center}
\end{table}

\par
\begin{figure}[!htp]
%  \centering
  \begin{minipage}[t]{0.51\textwidth}
  \includegraphics[width=\textwidth]{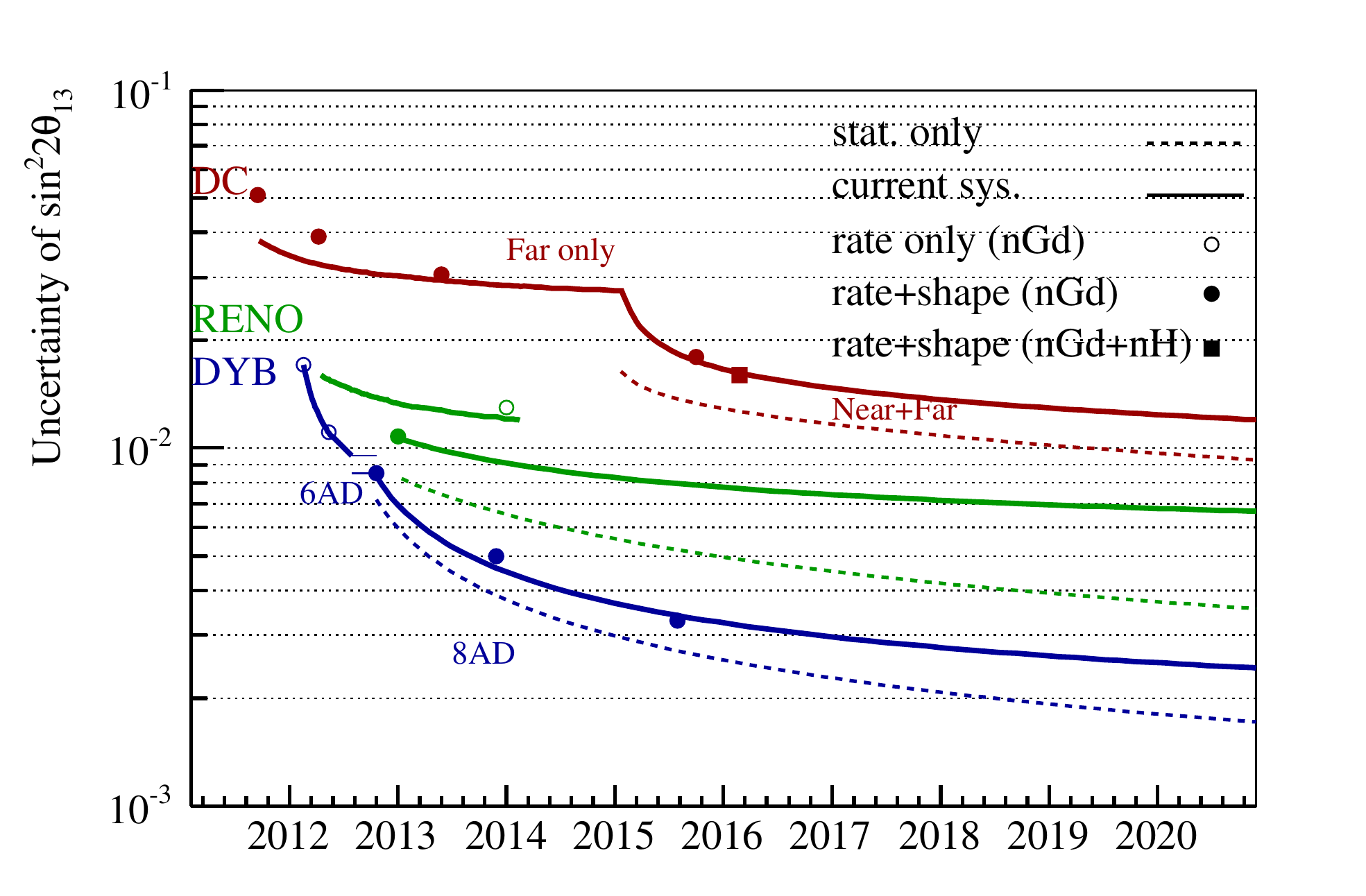}
  \end{minipage}
  \begin{minipage}[t]{0.51\textwidth}
  \includegraphics[width=\textwidth]{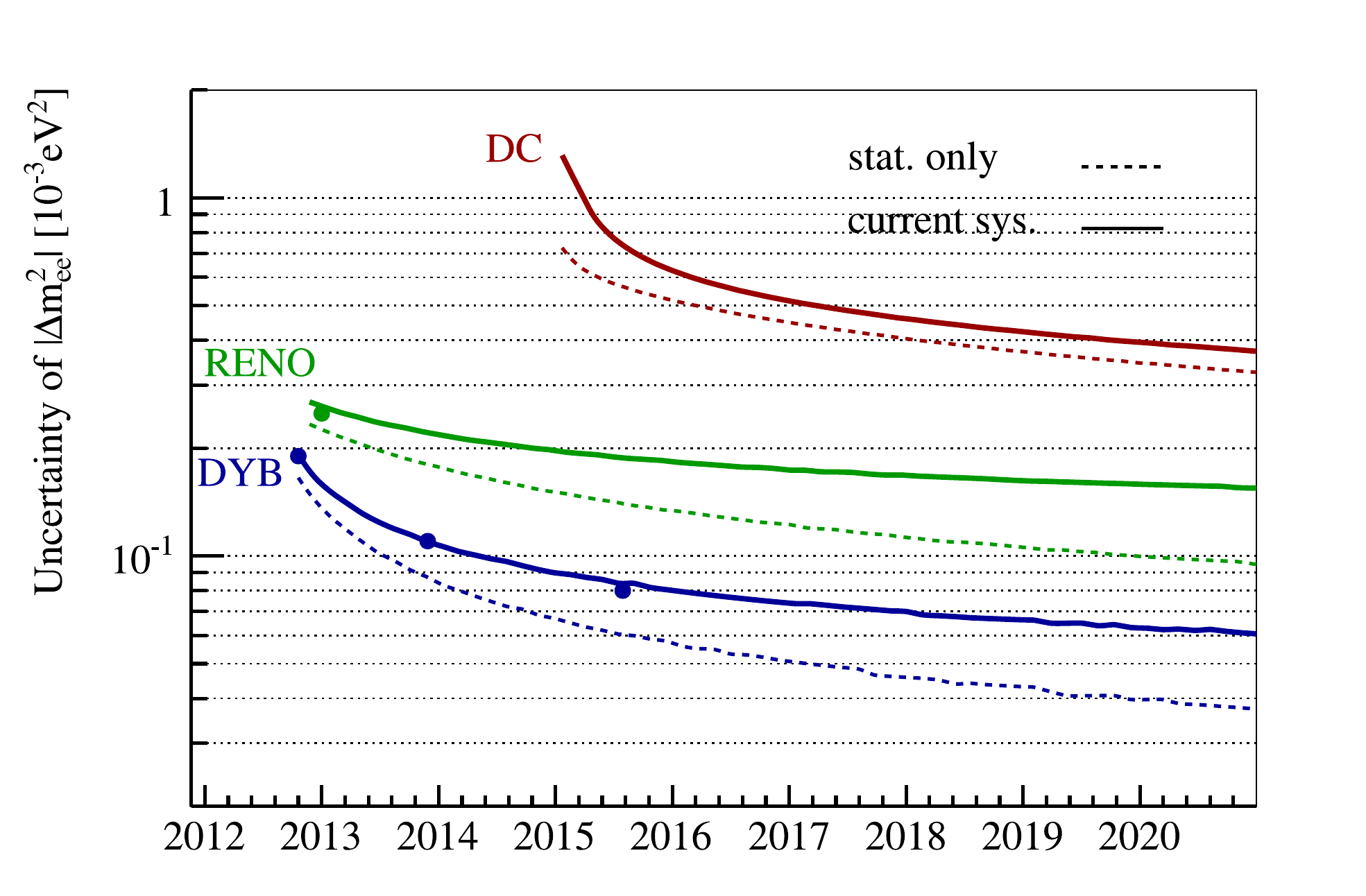}
  \end{minipage}
  \caption{Projected precision of $\sin^22\theta_{13}$ and $|\Delta m^2_{ee}|$ for the three experiments. The solid lines present the precision estimated with current systematics and the dashed lines show the statistical limit with zero systematic uncertainty. The points on the curves show the precision of published results for Daya Bay~\cite{Dayabay,An:2013uza,Dayabay14,An:2015rpe}, Double Chooz~\cite{DChooz,DChooz14,Abe:2014bwa,Abe:2012tg} and RENO~\cite{Reno,RENO:2015ksa}. The hollow markers refer to the rate-only analysis and the solid markers refer to the rate and shape analysis.}
  \label{fig:futurePrec}
\end{figure}

Around 2008, the global fit~\cite{Fogli:2008jx,Balantekin:2008zm,Schwetz:2008er} hinted the possibility of non-zero $\theta_{13}$. After the indications from T2K~\cite{Abe:2011sj}, MINOS~\cite{Adamson:2011qu}, and Double Chooz~\cite{DChooz} in 2011, Daya Bay discovered the oscillation due to $\theta_{13}$ at 5.2$\sigma$~\cite{Dayabay} in 2012, which was soon confirmed by RENO~\cite{Reno}. The most precise measurement shows that $\sin^22\theta_{13}\sim0.084$ with a precision of 4\% and $|\Delta m^2_{ee}|\sim2.50\times10^{-3}$ eV$^2$ with a precision of 3.3\%~\cite{An:2016ses}. The evolution of the measurements of these two fundamental parameters is shown in Figure~\ref{fig:futurePrec}. The Daya Bay experiment is expected to operate until 2020; by then, the precision will be $\sim\,$3\% for both $\sin^22\theta_{13}$ and $|\Delta m^2_{ee}|$~\cite{Cao:2016vwh}. Daya Bay has also significantly extended the exclusion area of the sterile neutrino searches~\cite{DayaBaySterile,An:2016luf,Adamson:2016jku} and obtained the most precise reactor \nuebar spectrum~\cite{An:2015nua,An:2016srz}.

\subsection{JUNO}
\label{sec:juno}

The large $\theta_{13}$ opened the gateway to determine the neutrino mass ordering (NMO) and to measure leptonic CP violation. In the early 2000s, the combined limit from Palo Verde, CHOOZ and Super Kamiokande was $\sin^22\theta_{13}<0.12$ (90\% C.L., if $\Delta m^2_{23}=0.0024$ eV$^2$), and the common intuition in the neutrino community was that $\theta_{13}$ would be tiny. Thus, it was generally believed that the NMO should be determined at long baseline experiments using accelerator neutrino beams. The possibility of distinguishing the NMO by exploring the effect of interference between the atmospheric- and solar- $\Delta m^2$ driven oscillations was first discussed, together with exploring the high-LMA MSW solution of the solar neutrino problem ~\cite{Petcov:2001sy,Choubey:2003qx}. Then the capability of determining the NMO by reactor neutrino experiments at a intermediate baseline was investigated by using a Fourier transform to the the $L/E$ spectrum~\cite{Learned:2006wy}. A more advanced Fourier analysis to the $L/E$ spectrum was performed~\cite{Zhan-PRD08,Zhan-PRD09}, and worked out the initial experimental requirements of using reactor antineutrinos to determine the NMO: a large LS detector with 3\%$/\sqrt{E}$ energy located at a baseline around 58 km from a powerful reactor complex (e.g, 24 GW). Later a standard $\chi^2$ analysis~\cite{Li-PRD13} demonstrated the unambiguous determination of the NMO using reactors, taking into account the impacts from real spatial distribution of reactor cores, residual energy non-linearity and the precision of the effective mass-squared difference $\Delta m^2_{\mu\mu}$.

\begin{figure}[!thb]
\begin{centering}
\includegraphics[width=0.6\textwidth]{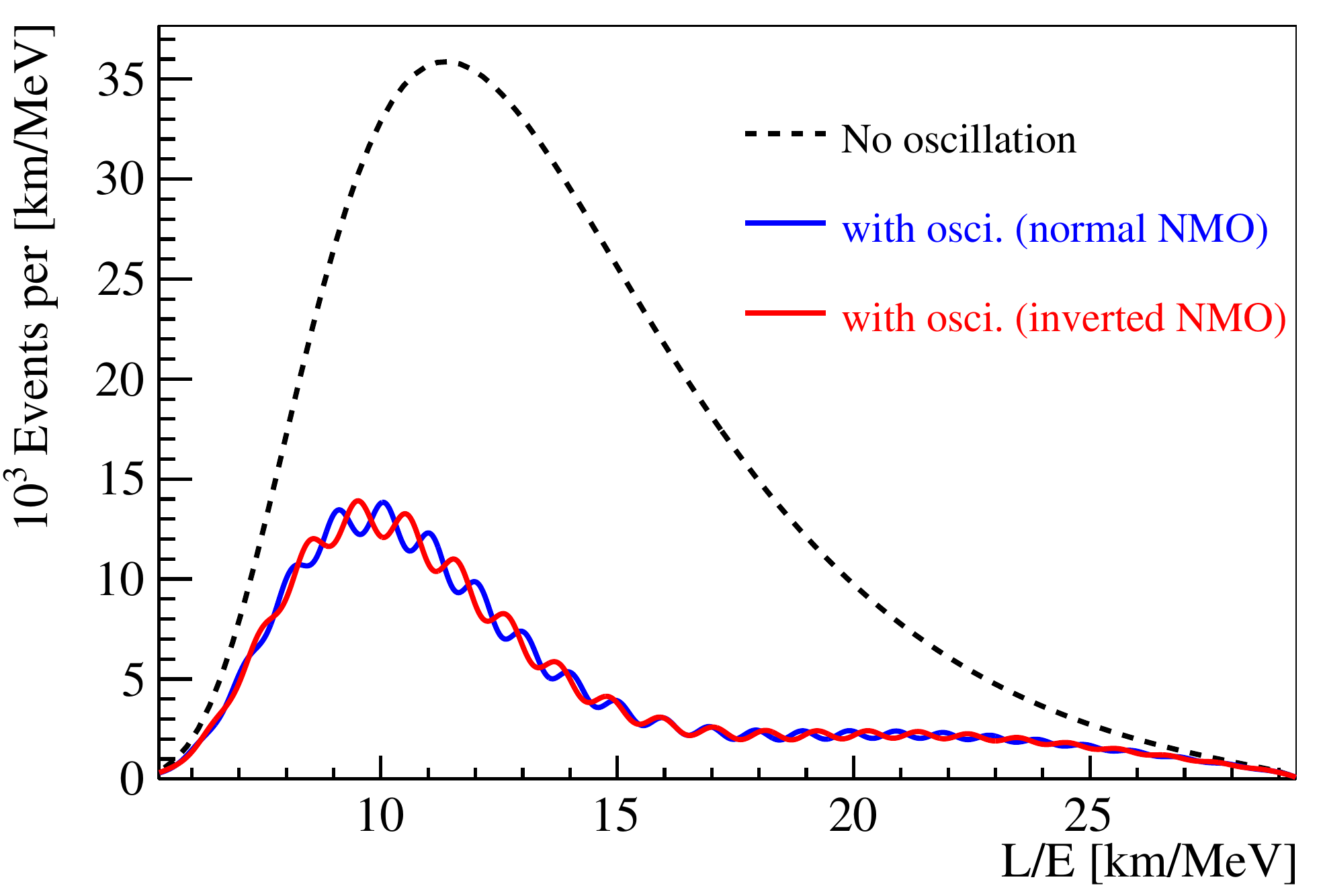}
\par\end{centering}
\caption{\label{fig:neuSpec} Reactor \nuebar spectrum at JUNO for no oscillations (dashed) and for different NMOs: $\Delta m^2_{32}=+2.4\times10^3$eV$^2$ (blue) for normal NMO and $\Delta m^2_{32}=-2.4\times10^3$eV$^2$ (red) for inverted NMO. The total exposure is normalized to 2,700 GW$\cdot$kton$\cdot$yr. No energy smearing is included to demonstrate the principle.}
\end{figure}

The Jiangmen Underground Neutrino Observatory (JUNO), proposed in 2008 and approved in 2013, is under design and construction, with the primary goal to determine the NMO~\cite{Djurcic:2015vqa,An:2015jdp}. The expected reactor \nuebar spectra at JUNO for different NMOs is shown in Figure~\ref{fig:neuSpec}. JUNO is located in Kaiping, a small town near Guangzhou in South China, $\sim\,$53 km from the Yangjiang and Taishan nuclear power plants. A total power of 26.6 GW will be available by 2020 when JUNO is scheduled to start data taking. The designed energy resolution of 3\%/$\sqrt{E \text{(MeV)}}$ is expected to be achievable with the high photocathode coverage and highly transparent liquids. The details about detector design and R\&D results will be described in Section~\ref{sec:futureExp}.

Detailed studies about the sensitivity of determining the NMO~\cite{An:2015jdp} demonstrated that a median sensitivity of $\sim3\sigma$ can be achieved with the reasonable assumption of the systematics and six years of running. Additional sensitivity can be gained by including precision measurement of $|\Delta m^2_{\mu\mu}|$ from future long baseline $\nu_\mu$ ($\overline\nu_\mu$) disappearance measurements. A confidence level of 3.7$\sigma$ or 4.4$\sigma$ can be obtained, for the $|\Delta m^2_{\mu\mu}|$ uncertainty of 1.5\% or 1\%. Given the 53 km baseline, the terrestrial matter effects are not negligible for JUNO, and such effects were evaluated to reduce the sensitivity of the NMO measurement by $\Delta\chi^2_{MO}\simeq0.6$ ~\cite{Li:2016txk}.

The precise prediction of the antineutrino spectrum for JUNO is needed. Unlike Daya Bay, it is unpractical to build a near detector for JUNO. However, Daya Bay can be treated as a virtual near site, based on that JUNO has similar reactor cores as Daya Bay, thus a virtual relative measurement can be taken, see discussions in Section~\ref{sec:fissionSpec}.

Precision measurement of the oscillation parameters and test of the standard three-neutrino framework are another important goals of the JUNO experiment. JUNO would be the first experiment to simultaneously observe the neutrino oscillation driven by both atmospheric and solar neutrino mass-squared differences. The oscillation parameters $\sin^2\theta_{12}$, $\Delta m^2_{21}$ and $\Delta m^2_{ee}$ can be measured to the world-leading precision of 0.7\%, 0.6\% and 0.5\%, respectively. The superior detector properties of JUNO also provide great opportunities in studying neutrinos from Supernova, the Earth's interior and the Sun, atmospheric neutrinos, sterile neutrinos, Nucleon decays, neutrinos from dark matter and other exotic searches. Details were discussed in~\cite{An:2015jdp}.

%% file: section_3.tex
%!TEX root = reactor_AR_main.tex
\section{REACTOR NEUTRINO FLUX AND SPECTRA}
\label{sec:flux}

\subsection{Predicting the flux and spectrum}

\par
Reactor neutrino flux and spectrum can be predicted with the fission rate of each isotope in the reactor core and the corresponding cumulative neutrino spectrum per fission. The fission rates can be estimated with the core simulation and the thermal power measurements. The neutrino spectrum per fission of each isotope can be determined by inversion of the measured $\beta$ spectra of fissioning with an uncertainty of 2-5\%~\cite{VonFeilitzsch:1982jw, Schreckenbach:1985ep, Hahn:1989zr, Huber:2011wv, Mueller:2011nm, Hayes:2016qnu}, or by summation of thousands of decay branches of the fission products with information in the nuclear database with an uncertainty of $\sim\,$10\%~\cite{Vogel:1980bk, Hayes:2016qnu}.

\par
Fission rates in a reactor are proportional to the thermal power during real operation. Instead of fission rates, normally we use fission fractions in the core simulation, which is the ratio of the fission rate of an isotope over the total rate. The reactor neutrino spectrum can be calculated as
\begin{eqnarray}
\label{eq:nuflux}
\Phi(E_{\nu})=\frac{W_{th}}{\sum_i f_i e_i}\cdot\sum_i f_i \cdot S_{i}(E_{\nu}),
\end{eqnarray}
Where $E_\nu$ is the neutrino energy, and $f_i$, $e_i$ and $S_i(E_{\nu})$ are the fission fraction, thermal energy released in each fission, and neutrino spectrum per fission for the $i$-th isotope, respectively.

\par
Fresh fuel contains only uranium. Plutonium is gradually generated via the neutron capture of $^{238}$U and the subsequent evolution. Generally a core refuels every 12-18 months, and replaces 1/4 to 1/3 fuel assemblies each time. There are many commercial or publicly available software to simulate the fuel evolution in the core. Uncertainties of the simulation, typically $\sim\,$5\% in terms of the fission fraction of each isotope, were estimated by comparing the simulated fuel composition at different burnup with isotopic analyses of spent fuel samples taken from the reactors. The thermal power $W_{th}$ can be measured to sub-percent level online. Although the fission fractions predicted by simulation carry large uncertainty, their contribution to the total uncertainty of the reactor neutrino flux is at sub-percent level, given the similarity among the four isotopes in the neutrino spectrum and energy released per fission.

\par
After refueling, the spent fuel taken out from the previous cycle are moved to a cooling pool adjacent to the core. The long-lived fission fragments will continue to decay and generate antineutrinos. Typically the spent fuel contributes about 0.3\% of the total antineutrinos when detecting via inverse $\beta$-decay reaction (IBD), depending on the storage and burnup information of the spent fuel.

\par
When evaluating the neutrino spectrum per fission by the conversion method, the non-equilibrium contribution should be taken into account. In ILL measurements, fissile samples were exposed to neutrons for one to two days. The beta decays from the long-lived fission fragments not reaching equilibrium, similar to those in the spent fuel, were missed in the ILL measurements. This contributes to 0.3-0.6\% of the total antineutrinos.

\subsection{Neutrino spectrum per fission}
\label{sec:fissionSpec}
\par
Neutrino spectrum per fission for each isotope, $S_{i}(E_{\nu})$, is the largest uncertainty in predicting the reactor neutrino flux. It could be calculated by superposing thousands of $\beta$-decays of the fission fragments. Such a first-principle calculation is challenging due to missing or inaccurate data even with modern nuclear databases. In general the uncertainty is $\sim\,$10\%~\cite{Vogel:1980bk, Hayes:2016qnu}. However, since it provides physical insights of the reactor neutrino flux and reactor physics, the evaluation keeps improving~\cite{Kopeikin:2003gu, Fallot:2012jv, Hayes:2013wra, Dwyer:2014eka}, especially after the discrepancy around 5 MeV was found.

\par
To improve on the purely {\it ab initio} method, several direct measurements were done at ILL in the 1980s to determine the neutrino fluxes and energy spectra of the thermal fissile isotopes $^{235}$U, $^{239}$Pu and $^{241}$Pu~\cite{VonFeilitzsch:1982jw, Schreckenbach:1985ep, Hahn:1989zr}. In these measurements, sample foils were placed into a reactor and exposed to neutrons for one or two days. A high precision electron spectrometer recorded the emitted $\beta$ spectra, which were then inverted to the antineutrino spectra by fitting the observed $\beta$ spectra to a set of 30 virtual $\beta$-branches. With the Q-value and the branching ratios of the virtual $\beta$-branches, the corresponding neutrino spectra could be computed out. The uncertainty of the antineutrino spectrum by this conversion process was estimated to be 2.7\%. These experiments did not perform similar measurements for $^{238}$U, which only fissions with fast neutrons. Combined with the {\it ab initio} calculations for $^{238}$U by Vogel~\cite{Vogel:1980bk}, with uncertainties of $<$10\%, the obtained reactor neutrino spectra are referred to as the ILL+Vogel model. Since $^{238}$U only contributes to $\sim\,$8\% of the total reactor antineutrino flux, the error introduced to the total flux is less than 1\%.

\par
The inversion method was later improved by Huber~\cite{Huber:2011wv}, in which the ILL data was re-analyzed with higher order corrections in the $\beta$ decay spectrum taken into account. The spectrum of $^{238}$U was also updated with an {\it ab initio} calculation by Mueller {\it et al.}~\cite{Mueller:2011nm}. Comparing to the ILL+Vogel model, the Huber+Mueller model shows a 3.5\% increase in total flux and a small excess in the high energy part of the spectra. The flux uncertainty is reduced to 2.4\%. The upward shift in the total flux introduces tension with short baseline reactor neutrino experiments (for a review see Ref.~\cite{Vogel:2015wua}), which is called the Reactor Neutrino Anomaly~\cite{Mention:2011rk}. A measurement of the $^{238}$U $\beta$ spectrum was performed and the corresponding antineutrino spectrum was determined in Ref.~\cite{Haag:2013raa}, which is different from the {\it ab initio} calculation by only 0.2\% in integrated flux.

\begin{figure}[thb]
\begin{centering}
\includegraphics[width=0.7\textwidth]{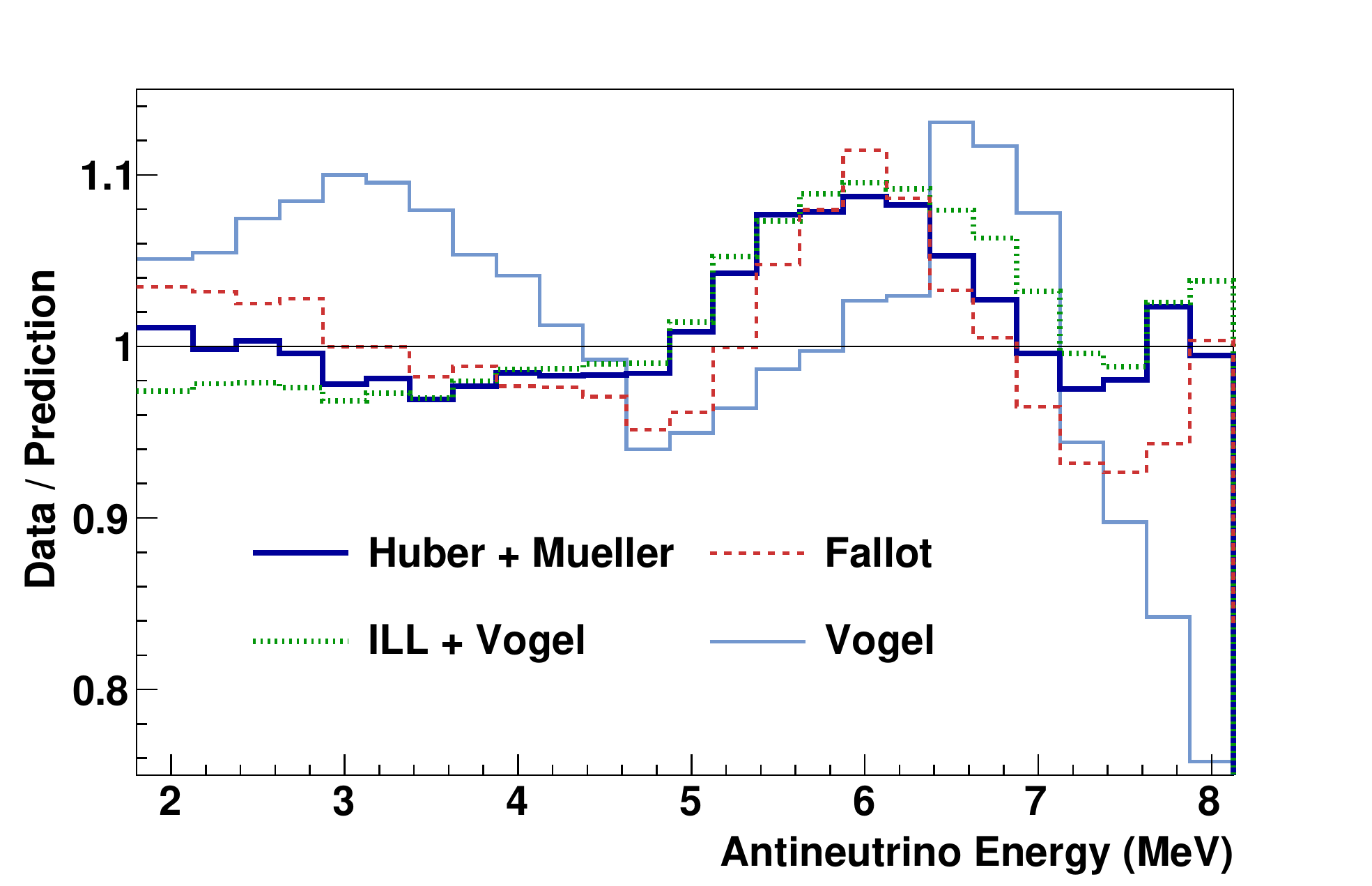}
\par\end{centering}
\caption{\label{fig:rcs} Comparison of two {\it ab initio} calculations (Vogel~\cite{Vogel:1980bk} and Fallot~\cite{Fallot:2012jv}), two inversion models (ILL+Vogel and Huber+Mueller), and spectrum measured by Daya Bay, shown as ratios of the Daya Bay measurement over predictions.}
\end{figure}

\par
Recently, Daya Bay~\cite{An:2015nua}, Double Chooz~\cite{Abe:2014bwa}, and RENO~\cite{RENO:2015ksa} have found significant local inconsistency (at $4\sim6$ MeV) between the measured and the predicted reactor neutrino spectrum, no matter using the ILL$+$Vogel or the Huber$+$Mueller flux model. The largest deviation reaches $\sim\,$10\%, significantly larger than the expected uncertainty 2-3\%. A comparison of the {\it ab initio} calculations and inversion models with respect to the Daya Bay measured neutrino spectrum~\cite{An:2016srz} is shown in Figure.~\ref{fig:rcs}.

\par
The uncertainty of the Daya Bay measured spectrum ranges from 3\% to 9\% in 1.8-7.6 MeV. The dominant uncertainty is from the energy non-linearity of the detector while the statistical uncertainty is at percent level (see Fig.~29 in Ref.~\cite{An:2016srz}). With on-going dedicated detector studies and more data, the precision will be improved. When predicting flux and spectrum for future reactor neutrino experiments like JUNO from the Daya Bay measurements, the uncertainty could be much smaller than the above absolute uncertainty since the relative energy non-linearity between the Daya Bay and JUNO detectors could be measured much easier than the absolute energy non-linearity. The Daya Bay measured spectrum is valid only for given fission fractions. Fortunately the average fission fractions of light water reactors are similar when integrating over a certain time. The small deviation from the Daya Bay average fission fraction can be complemented with the inversion models. There are many short-baseline experiments planned or undergoing with different detector technologies. Model independent predictions based on the new precision measurements can avoid the bias in the inversion models, and might be able to improve the precision to 1\%.

\subsection{Uncertainties}

\par
The uncertainties of the integrated reactor antineutrino flux is listed in Table~\ref{tab:ReactorUncertainty}.
The thermal power uncertainty is typically 0.5\%, which can be found in the power plant documents. It should be noted that generally the uncertainty provided by the power plant is at 95\% confidence level. The fission fractions are correlated in the fuel evolution and can be studied with the core simulation~\cite{Ma:2014bpa}. When constrained by the total thermal power and taking the correlation into account, the typical 5\% fission fraction uncertainty of a single isotope converts to 0.6\% uncertainty in total flux. In the table, uncertainty of the neutrino spectrum per fission is taken from the Huber+Mueller model which is underestimated as found by recent measurements. The uncertainty of the inversion methods were reevaluated to be 5\%~\cite{Hayes:2016qnu}. When detecting via inverse $\beta$ decay, the uncertainty of calculated cross section relates to the neutron lifetime, which is $0.12\%$.

\begin{table}[!htb]
\begin{center}
\caption{Typical uncertainties of the integrated reactor \nuebar flux, taken from Ref.~\cite{An:2016srz}.}
\begin{tabular}{c|c}
\hline
& uncertainty   \\\hline
Power & 0.5\%   \\\hline
Energy per fission & 0.2\%  \\\hline
Fission fraction & 0.6\%  \\\hline
Neutrino spectrum per fission & 2.4\%  \\\hline
Spent fuel & 0.3\%  \\\hline
Non-equilibrium & 0.2\% \\\hline
IBD cross section & 0.12\%  \\ \hline
\end{tabular}
\label{tab:ReactorUncertainty}
\end{center}
\end{table}

\par
For a single detector experiment, all uncertainties are included. For a relative measurement with multiple detectors, uncertainties correlated among reactors will cancel out, including that of energy per fission and the IBD cross section. Uncorrelated uncertainty may still contribute, including that of power, fission fraction, neutrino spectrum per fission, spent fuel, and the non-equilibrium contributions.

%% file: section_4.tex
%!TEX root = reactor_AR_main.tex

\section{TOWARD A PRECISION MEASUREMENT}
\label{sec:expTech}

\par
%The reactor experiments before 2000 have reactor related uncertainties of 2$\sim\,$3\%, background related uncertainties of 1$\sim\,$3\%, and detector related uncertainties of 2$\sim\,$4\%. Current experiments Daya Bay, Double Chooz, and RENO have significantly reduced the systematic uncertainties, with the relative measurement approach and improved technologies. In the Daya Bay proposal~\cite{Guo:2007ug}, the projected detector uncertainty per module was 0.38\% and the targeted background uncertainty was 0.24\% (0.37\%) for near sites (Far site). Recently Daya Bay achieved 0.13\% in detector uncertainty~\cite{An:2016ses}, and $\sim\,$0.15\% in the uncertainty of total background to signal ratio. This signals the extremely high precision era of neutrino physics. The experimental technologies and detector designs that lead to this great precision is reviewed below. // this is the original version in the submitted file

Reactor experiments before 2000 had reactor-related uncertainties of $\sim$2-3\%, background-related uncertainties of $\sim$1-3\%, and detector-related uncertainties of $\sim$2-4\%. Mikaelyan and Sinev pointed out that, the systematic uncertainties can be greatly suppressed or totally eliminated, if performing a near-far relative measurement with two detectors positioned at different baselines (42). The near detector close to the reactor core is used to used to establish the flux and energy spectrum of the electron antineutrinos. This relaxes the requirement of knowing the details of the fission process and operational conditions of the reactor. In this way, the value of $\sin^22\theta_{13}$ can be measured by comparing the electron antineutrino flux and energy distribution observed with the far detector to those of the near detector, after scaling with the square of the baseline distance.

There are two approaches, namely the rate analysis and the shape analysis. The rate analysis is on the basis of the ratio of the electron antineutrino rates measured at two baselines. If the near and far detectors are made identical, the absolute values of detection efficiencies and target numbers are practically canceled, and only their small relative differences are to be considered. Then, the ratio, dependent only on the distances and the survival probabilities, is used to extract $\sin^22\theta_{13}$.
The shape analysis is based on the comparison of the shapes of the electron antineutrino spectra measured simultaneously in two detectors. Small deviations of the ratio from a constant value, are searched for the oscillation effects. The detailed knowledge of that constant term, related to the geometry, target numbers and efficiencies, is no longer needed. In practice, the idea of near-far relative measurement has to be extended to handle more complicated arrangements involving multiple reactors and multiple detectors. Furthermore, each source of uncertainty should be classified correctly into correlated and uncorrelated uncertainties.

The current experiments Daya Bay, Double Chooz, and RENO have significantly reduced the systematic uncertainties, with the relative measurement approach and improved technologies. In the Daya Bay proposal~\cite{Guo:2007ug}, the projected detector uncertainty per module was 0.38\% and the targeted background uncertainty was 0.24\% (0.37\%) for near sites (Far site). Recently Daya Bay achieved 0.13\% in detector uncertainty~\cite{An:2016ses}, and $\sim\,$0.15\% in the uncertainty of total background to signal ratio. This signals the extremely high precision era of neutrino physics. The experimental technologies and detector designs that lead to this great precision is reviewed below.

\subsection{Liquid Scintillator}
\label{sec:LS}

\par
Liquid scintillator has been exclusively used in reactor neutrino experiments. Gadolinium doped liquid scintillator (Gd-LS) was popularly used due to two advantages. First, Gadolinium isotopes have large cross sections of thermal neutron capture. Second, the high-energy $\gamma$ cascade ($\sim\,$8MeV) after n-capture provides distinguishable delayed signal from the natural radioactivity.

Palo Verde was successful in Gd-LS, which has an attenuation length of $\sim\,$11 m, and the average degradation was $\sim\,$1 mm/day over two years~\cite{Boehm:2000vp}. CHOOZ had a serious problem of Gd-LS aging. The reduced transparency of Gd-LS due to chemical instability forced a replacement after 4 months of operation~\cite{Apollonio:1997xe}. A new filling was done, but the LS transparency still showed a fast degradation over time, due to the oxidation by nitrate ions. The transparency degradation forced frequent checks to the attenuation length and resulted in the energy threshold varying with time.

Daya Bay and RENO chose LAB as the LS solvent, due to its long absorption length, good stability, high flash point, good chemical compatibility with acrylic and low toxicity. The attenuation length of the production Daya Bay Gd-LS was $\sim\,$15 m at 430 nm~\cite{Beriguete:2014gua}. Double Chooz chose PXE, and the measured attenuation length for Gd-LS was 7.8$\pm$0.5 m at 430 nm~\cite{Aberle:2011ar}.

Dissolving the inorganic Gd salt in the aromatic scintillator solvents is non-trivial, and formation of Gd organic complex is an effective way. Carboxylic acids, $\beta$-diketones and organophosphorous compounds were the commonly used organic ligands~\cite{Piepke:1999db,Lightfoot:2004dj,Danilov:2003,Yeh:2007zz} to form the Gd complex. Palo Verde successfully synthesized a Gd and EHA complex~\cite{Piepke:1999db}, but its solubility in LAB was relatively poor. Daya Bay chose TMHA as the complexing ligand to form an organo-complex (Gd-TMHA) with gadolinium chloride~\cite{Ding:2008zzb}, because of good solubility up to 10 g/L and good attenuation length. Double Chooz chose $\beta$-diketones for better stability of Gd-LS. The running reactor experiments for $\theta_{13}$ all used 2,5-diphenyloxazole (PPO) and 1,4-bis[2-methylstyryl]benzene (bis-MSB) as the primary fluor and wavelength shifter, respectively.

Raw material purification is necessary to remove the colored contaminants (iron and cobalt) and the natural radioactive U/Th isotopes. The presence of iron not only decreases optical transparency in the sensitive region of the PMT, but also degrades chemical stability of Gd-LS. The radiation from U/Th decay chains can form accidental coincidences mimicking \nuebar signal. Particularly, the emitted alpha particles can produce correlated background via $^{13}$C($\alpha$, n)$^{16}$O reactions (see Section~\ref{sec:radControl}). Daya Bay conducted purifications such as filtration after melting, distillation and recrystallization to purify PPO, and thin-film vaccum distillation to purify THMA~\cite{Beriguete:2014gua}. In addition, a co-precipitation approach was developed to remove U/Th from the raw materialGdCl$_3\cdot x$H$_2$O~\cite{Yeh:2010zz}. The long-lived radium in the U/Th chain cannot be removed by this approach but it will not complex with the ligand like Gd and U/Th. This approach both removed the radioactive isotopes and significantly improved the optical transparency~\cite{Beriguete:2014gua}.

An important source of non-linear and non-uniform energy responses in LS is due to complex absorption and re-emission processes for scintillation and Cerenkov photons. Precise simulation of the light propagation in LS is highly desired, and can consequently reduce the energy scale uncertainty. A generic optical model in Monte Carlo has been developed to describe the complete absorption and re-emission processes (see Section~\ref{sec:junoReso}), which is expected to significantly improve the understanding of the LS non-linearity.

\subsection{Segmented vs. Homogeneous Detector}

The shallow depth forced Palo Verde to end up with a unique segmented design for the purpose of suppressing fast neutron backgrounds. However, the segmented design resulted in no monochromatic peak, causing difficulties for understanding the energy resolution, efficiency and systematic errors. On the contrast, the homogeneous design of CHOOZ gave a full-absorption peak of 8 MeV $\gamma$s from the n-capture on Gd, leading to high efficiency and smaller uncertainty on neutron detection. Palo Verde had 12 tons of target and only $\sim\,$10\% efficiency, whereas CHOOZ had only 5 tons Gd-LS but $\sim\,$70\% efficiency.

Segmented detectors have larger edge effects. Palo Verde observed two time constants for the neutron capture due to the inhomogeneity of its target. The faster time constant (27.1 $\mu$s) was for the neutrons remaining in the Gd-LS target, and the slower constant (76.8 $\mu$s) was indicated by a GEANT3 simulation as due to the neutrons which enter the acrylic.

\subsection{Identical Detectors}

The reactor experiments for $\theta_{13}$ opened the high precision era of neutrino experiments. For example, the detector-related systematic uncertainty of Daya Bay has reached 0.13\%~\cite{An:2016ses}. With identical detectors at the near and far sites, all uncertainties correlated among detectors cancel out in the near-far relative measurement. The residual uncertainties come from the tiny differences among detectors. Therefore, the key is to make the detectors functionally identical, both by design and by careful fabrication. Innovative ideas were brought into the design based on past experiences:
\begin{itemize}
  \item Three-zone structure detector to well define the target mass, reduce uncertainty of the neutron-tagging efficiency, and shield the radioactivities.
  \item Cylindrical detector to reduce the fabrication difficulties
  \item Same batch of liquid scintillators for all detectors.
\end{itemize}

\par
A spherical detector has more uniform response than a cylindrical one. However, cylindrical acrylic vessels are much easier to be fabricated identically. For instance, the diameter of the inner acrylic vessel of Daya Bay is designed to be 3120 mm, with 5 mm tolerance. The as-built surveys show that the actual diameter have a variation of only 2 mm, corresponding to 0.17\% variation of volume. This is comparable to the target mass variation which is controlled to be 0.19\% by precise load cell measurement during filling.

\par
Both Daya Bay and Double Chooz chose a nested, three-zone cylindrical structure, which has been a major step forward from the CHOOZ detector design toward the high precision design of reactor $\theta_{13}$ experiments. The Gd-LS target is well defined by the rigid inner acrylic vessel. Target mass can be precisely measured during filling and monitoring the liquid level in the overflow tank during operation. The LS between the inner and outer acrylic vessels absorb $\gamma$ rays leaking out from the target. Without this gamma catcher layer, the characteristic 8 MeV signal of reactor neutrinos from neutron capture on gadolinium will have a large tail for events near the edge of the target. A cut on the tail to select neutrons, e.g. at 5 or 6 MeV, could introduce large uncertainty due the energy scale difference among detectors. A very thick gamma catcher may eliminate this uncertainty, while the actual thickness for the three experiments ranges from 42.5 cm to 70 cm, balancing with the statistical uncertainty (i.e. target mass). The buffer oil between the outer acrylic vessel and the outermost stainless steel vessel shields the radioactivity from the PMT glass and other detector materials.

\par
One possible large source of uncertainty is the hydrogen and gadolinium content in the liquid scintillator, which is 0.8\% for CHOOZ. The $\theta_{13}$ experiments mixed Gd-LS and LS for all detectors in ``one batch" to make the detectors identical. In Daya Bay, 185 ton Gd-LS was produced in 4-ton batches using the same batch of materials, and stored in five 40-ton acrylic tanks. Each tank has a built-in circulation system to mix the Gd-LS uniformly. 4 ton Gd-LS was taken from each of the five storage tanks to fill into one detector, which has 20 ton target mass, to ensure identical Gd-LS content for all eight detectors.

\par
Besides the designs common in all three experiments, Daya Bay detectors have several unique features, e.g. multiple detectors at one site and the movable detector design.

\par
\begin{figure}[!thb]
\begin{centering}
\includegraphics[width=0.9\textwidth]{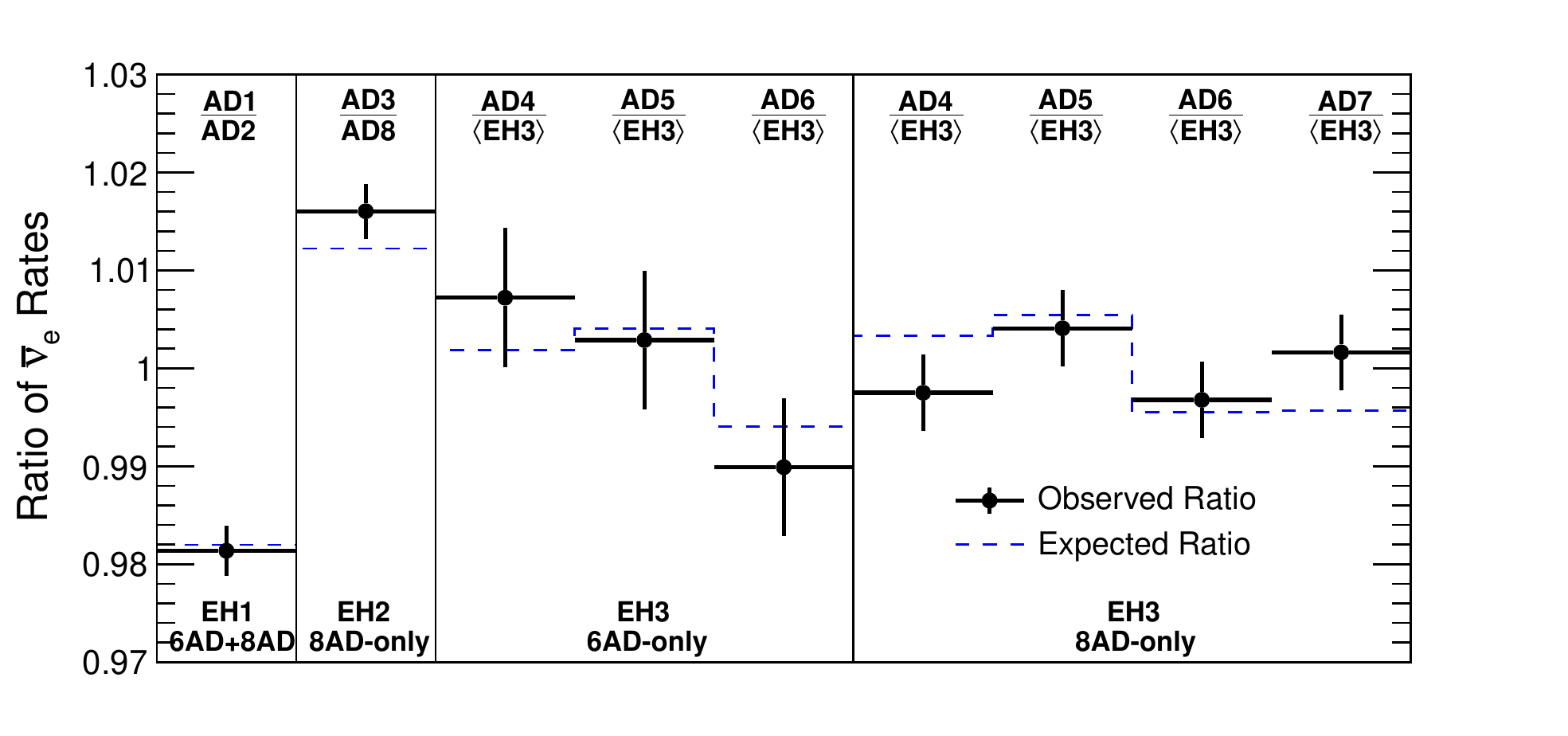}\label{fig:sidebyside}
\par\end{centering}
\caption{Side-by-side comparison of Daya Bay Detectors~\cite{An:2016ses}: ratios of the \nuebar interaction rates observed by detectors within the same experimental halls of the Daya Bay experiment.}
\end{figure}

Daya Bay has two detectors at each of the two near sites and four at the far site. Multiple detectors at the same site could cross-check detector systematics and statistically reduce the uncertainties. The Daya Bay detector is designed to have 0.38\% uncertainty for each, and turns out to be 0.2\% for early analyses and 0.13\% for the latest analysis. Such small uncertainty needs convincing demonstration. The side-by-side comparison ultimately demonstrates the small systematics~\cite{DayaBay:2012aa,An:2016ses}, shown in Figure~\ref{fig:sidebyside}. With this unique approach, the total systematic uncertainty actually can be ``measured" by comparing the expected ratio of \nuebar events in the side-by-side detectors with the measured ratio.

\par
Uncorrelated uncertainties by definition would be reduced by a factor of $1/\sqrt{N}$, where $N$ is the number of detectors. Ideally, making infinite number of identical detectors would suppress the uncorrelated uncertainties to zero. In reality, each detector needs a gamma catcher and a buffer layer of certain thickness to shield the target. The total target mass is determined by the requirement of \nuebar statistics. It is not cost-effective to have too many detectors. The eight detector configuration is almost optimal for Daya Bay.

\par
The Daya Bay detectors are designed to be movable. In case the detector uncertainty turns out to be large, swapping the near and far detectors could clarify the measurements. Since the actual uncertainty is much smaller than design, swapping is not necessary now. The movable design also enables the assembly and the LS filling, done with a specially designed 20-ton filling tank equipped with load cell, of all detectors done at the same site and with the same equipment, helping to make the detectors identical.

\subsection{Reflector in the detector}

\par
Reflective panels were first adopted by Daya Bay, which are put at the top and the bottom of the detector to increase light collection efficiency. Similar energy resolution is gained with only half of PMTs, thus greatly reduced the cost. There were worries that the light reflection may greatly complicate the understanding of the detector, resulting worse systematic uncertainties. Extensive Monte Carlo studies and a prototype experiment showed that a good agreement can be obtained~\cite{Wen:2011zzb}. A flat reflective panel could be fabricated as expected while a cylindrical side wall may easily deviate from the ideal shape. Therefore, the reflectors are only installed at the top and the bottom, and the cylinder wall is equipped with black acrylic sheets at the equator of the PMTs. The reflective panels were fabricated with great care~\cite{An:2015qga} and performed very well. The minor disadvantages include that the timing calibration with LED is more difficult, and the muon reconstruction is hard for those which penetrate the top and bottom reflectors.

\subsection{Energy nonlinearity}
\label{sec:nonEL}

Measurement of the reactor neutrino spectrum relies on the interpretation of the observed prompt energy spectrum. The energy response in the antineutrino detector is nonlinear and particle-dependent due to scintillator and possible electronics effects. It is not convenient to calibrate the nonlinearity with positrons directly. Instead, the nonlinearity normally needs to be deduced from calibration data or physics data with $\gamma$ and $\beta$. With extensive studies, it has been determined to 1-2\% uncertainty in the reactor neutrino energy range in the reactor $\theta_{13}$ experiments, and JUNO aims at $<\,$1\% uncertainty for the nonlinearity.

\par
The front-end electronics (FEE) of Daya Bay measures the integrated charge of PMTs in a fixed time window with a peak-finding algorithm~\cite{Guo:2007ug}. The charge integration introduce an electronics nonlinearity due to the loss of the slow scintillation light. The electronics nonlinearity interplays with the scintillator nonlinearity when calibrated with $\gamma$ or $\beta$ data. As for Double Chooz, the electronics nonlinearity was due to insufficient ADC precision that could bias the baseline estimation within $\pm$1 ADC unit, resulting in a charge-dependent gain nonlinearity especially below a few photoelectrons~\cite{Abe:2014bwa}. Since the spectrum measurement is critical for JUNO, all 20-in PMTs will be read out with 12bit Flash ADCs. In principle the nonlinearity from the electronics should be negligible.

\par
The scintillator nonlinearity is caused by the quenching effect described by Birks' law~\cite{Birks:1951boa}, as well as absorption and re-emission of Cerenkov photons. Since $\gamma$s deposit energy via Compton scattering, the nonlinearity of $\gamma$ is essentially a convolution of the energy distribution of the primary Compton electrons and the nonlinearity response of mono-energetic electrons. Daya Bay constructed an empirical model~\cite{Dayabay14} to describe the scintillator nonlinearity for electrons, then the relationship from the $e^-$ nonlinearity to the response for $\gamma$ and $e^+$ could be derived by Monte Carlo simulation. The energy model was determined by a fit to monoenergetic $\gamma$ lines from various radioactive sources and the $\beta$ spectrum from the cosmogenic \Btwelve events. This model was validated by 1) calibration radioactive sources ($^{68}$Ge, $^{60}$Co, $^{137}$Cs, $^{54}$Mn, $^{40}$K, n capture on H, C, and Fe from $^{241}$Am-$^{13}$C, $^{241}$Am-$^{9}$Be and Pu-$^{13}$C); 2) $\gamma$ peaks or $\gamma+\beta$ from natural radioactivity ($^{40}$K, $^{208}$Tl, $^{214}$Bi-$^{214}$Po-$^{210}$Pb cascade in U chain, and $^{212}$Bi-$^{212}$Po-$^{208}$Tl cascade in Th chain), 3) table-top measurements~\cite{Zhang:2014iza} via Compton scattering.

A 1 GHz Flash ADC system has been installed on one of the Daya Bay detectors to measure the electronics nonlinearity. Dedicated calibration has been done and LS optical property studies will be done soon. With these studies, it is expected to improve the uncertainty of the energy nonlinearity from 1-2\% to below 1\%.

\subsection{Radioactivity Control}
\label{sec:radControl}
\par
Ambient radioactivity can be easily shielded by immersing the neutrino detector into a water pool or by surrounding it with other clean material, while the radioactivity from the detector itself needs to be carefully controlled. Radioactivity control requirement for each detector component or material should be specified during design according to the difficulties of material selection or purification and the budget allocation of the total singles rate of the detector. Radio-assay should be done for all components that may have a large contribution. Proper shielding may be needed for some components.

\par
Stainless steel is commonly used for the outermost detector vessels and supporting structures. However, the typical radioactivity of commercial stainless steel is 0.1-10 Bq/kg, particularly with high activity of $^{60}$Co, if the raw material includes scrap steel. The straightforward solution for low radioactivity stainless steel is to use low-radioactivity iron ore and never use the scrap steel. Daya Bay successfully used this strategy to produce all 260 tons of low radioactivity 304L stainless steel in one batch.

\par
PMT glass generally carry high radioactivity and is always a major source of singles events in a LS detector, even when made with special recipe with low potassium content. It is important to procure low background PMTs. A buffer layer is needed to shield PMTs, as well as other supporting material, from the LS target.

\par
Acrylic normally carries little radioactivity, at several or tens ppt level. No special treatment was done for the $\theta_{13}$ experiments except the cleaning to remove dust. However, it will be important to JUNO, since the neutrino event rate will be much lower at 53 km baseline and the total mass of acrylic vessel ($\sim\,$600 ton) is huge. The production line for the acrylic sheet has been improved to distill the monomer and completely isolate the acrylic from air throughout the production. Special procedure has been established for the shaping, shipping, and bonding, including isolation from air, protection with Nitrogen, and cover the sheet surface with plastic film. The goal is to control the radioactivity of the acrylic vessel to 1 ppt in U/Th.

\par
Removal of alpha emitters from LS raw materials would significantly reduce the $^{13}$C$(\alpha, n)^{16}$O reactions in LS, which can mimic an IBD signature: the prompt signal is caused by the neutron elastic scattering on proton, or inelastic scattering on $^{12}$C, or the de-excitation $\gamma$s from the $^{16}$O excited states; the delayed signal is the neutron capture after thermalization. This background was overlooked in KamLAND until its second oscillation analysis~\cite{Araki:2004mb}, because the visible energy of $\alpha$ particle from $^{210}$Po was quenched below threshold in its first analysis. Eventually the ($\alpha,n$) reaction was the dominant background in KamLAND. The total $^{13}$C$(\alpha, n)^{16}$O reaction rate can be estimated by using the reaction cross section from nuclear databases and the simulated discrete energy stopping power of alpha particles.

\par
The dominant alpha sources in Gd-LS were found to be $^{238}$U, $^{232}$Th and $^{227}$Ac actinide decay chains identified via polonium cascade decays: $^{214}$Bi-$^{214}$Po-$^{210}$Pb in the U chain, $^{212}$Bi-$^{212}$Po-$^{208}$Pb in the Th chain, and $^{219}$Rn-$^{215}$Po-$^{211}$Pb in the Ac chain (daughter of $^{235}$U). Daya Bay measured 0.45 mBq/ton $^{238}$U, 8 mBq/ton $^{232}$Th and 10 mBq/ton $^{227}$Ac in Gd-LS, respectively. The eventual ($\alpha,n$) background was negligible in Daya Bay, thanks to the LS purification process described in Section~\ref{sec:LS}. The natural abundance ratio of $^{238}$U to $^{235}$U is $\sim\,$22, whereas for Daya Bay Gd-LS it was measured to be $\sim\,$0.05, indicating a purification of at least 400 times.

In underground laboratories, Radon generated inside the rock can penetrate from rock cracks, and be released in the air. Without proper protection, Radon can lead to a serious background. In general, the detector and the storage and piping system of the LS and water should be air-tight and protected with Nitrogen cover when necessary. JUNO has very high requirements on Radon. The activity in the water pool should be $<$ 0.2 Bq/m$^3$. Besides the protection, the water flow will be controlled to reduce the background rate in the LS detector.

\subsection{Fast neutron background}

Muon-induced neutrons are important backgrounds in reactor neutrino experiments. The primary production processes are the $\mu$-nuclei interactions via virtual $\gamma$s (referred as ``muon spallation"). The huge difference between early theoretical calculations and experimental results have been understood to be the cause of subsequent electromagnetic and hadronic showers induced by gammas, pions and neutrons~\cite{Wang:2001fq}. Secondary processes such as photonuclear interactions, giant resonance decays, evaporation etc. also play important roles. FLUKA~\cite{Fasso:1997cb,Ferrari:2005zk}, the only tool with the correct cross sections for these processes in early 2000, was used to quantitatively study the dependence of the spallation neutron yield on depth and parent muon energy~\cite{Wang:2001fq}, and an empirical parametrization of yield and angular distributions was derived. The FLUKA results were confirmed later by a Geant4 simulation~\cite{Marino:2007ti}, where new hadronic processes were included and secondary neutrons were properly handled.

\par
Energetic, or {\it fast}, neutrons can mimic an IBD reaction: the recoil proton generates the prompt signal and the capture of the thermalized neutron provides the delayed signal. Double neutron captures can also mimic an IBD signal. Palo Verde suffered from fast neutrons, which was not easy to model at that time, and the background estimation from simulation had a large uncertainty. A $swap$ method~\cite{Wang:2000kt} was invented to suppress the uncertainty from neutron-induced backgrounds, based on the fact that the \nuebar signals and neutron backgrounds showed different symmetry behavior if the selection cuts were reversed by imposing the neutron cuts on the prompt signal and the positron cuts on the delayed signal. It reduced the measurement errors on the \nuebar flux from $\sim\,$20\% to $\sim\,$10\%.

\par
The best shielding to attenuate fast neutrons is water, rather than sand or steel, because it is rich in protons, radio-pure, and cost-effective. The water buffer can provide the best tagging of muons if equipped with photomultipliers acting as a water Cerenkov detector. The lateral distance that fast neutrons can travel in water from the muon track, is approximately exponential with an average distance of 0.7$\sim\,$0.8 m. The water pool of Daya Bay was designed to provide at least 2.5 m thick water shielding in all directions, whereas the inner veto system of Double Chooz has 50-cm thick liquid scintillator, and the veto detector of RENO has 1.5-m thick water.

\par
The broad continuum spectrum of fast neutrons can be revealed by expanding the selection of the \nuebar candidates with relaxed prompt energy cuts. No physical mechanism guarantees a particular shape, e.g, flat or linear, for the fast neutron spectra, because such backgrounds highly depend on the design and performance of the muon veto system, as well as the veto strategy in analysis. The three $\theta_{13}$ experiments observed different shapes of fast neutrons.  In Daya Bay, although the simulation supported the validity of a linear extrapolation of this background into the \nuebar signal region, more robust data-driven approaches were used to estimate this background and uncertainties~\cite{An:2016ses}. With tagged and untagged fast neutron samples, and their comparison with MC simulation, the fast neutron backgrounds can be well understood within $\sim\,$15\% uncertainty.

\subsection{Cosmogenic $^{9}$Li/$^{8}$He background}

The cosmogenic $\beta$-n emitters, \Linine/\Heeight, are the most serious correlated backgrounds in reactor neutrino experiments. Such background was neglected by Palo Verde and CHOOZ, but was dominant for KamLAND. Before KamLAND was built, the cross sections of muon-induced radioactive isotopes were measured with the SPS muon beam on $^{12}$C target at CERN~\cite{Hagner:2000xb}. Later KamLAND data demonstrated that \Linine/\Heeight are predominantly produced by energetic muon showers in the LS~\cite{KamLAND-spall}.

\begin{figure}[thb]
\begin{minipage}[t]{0.5\textwidth}
\includegraphics[width=\textwidth]{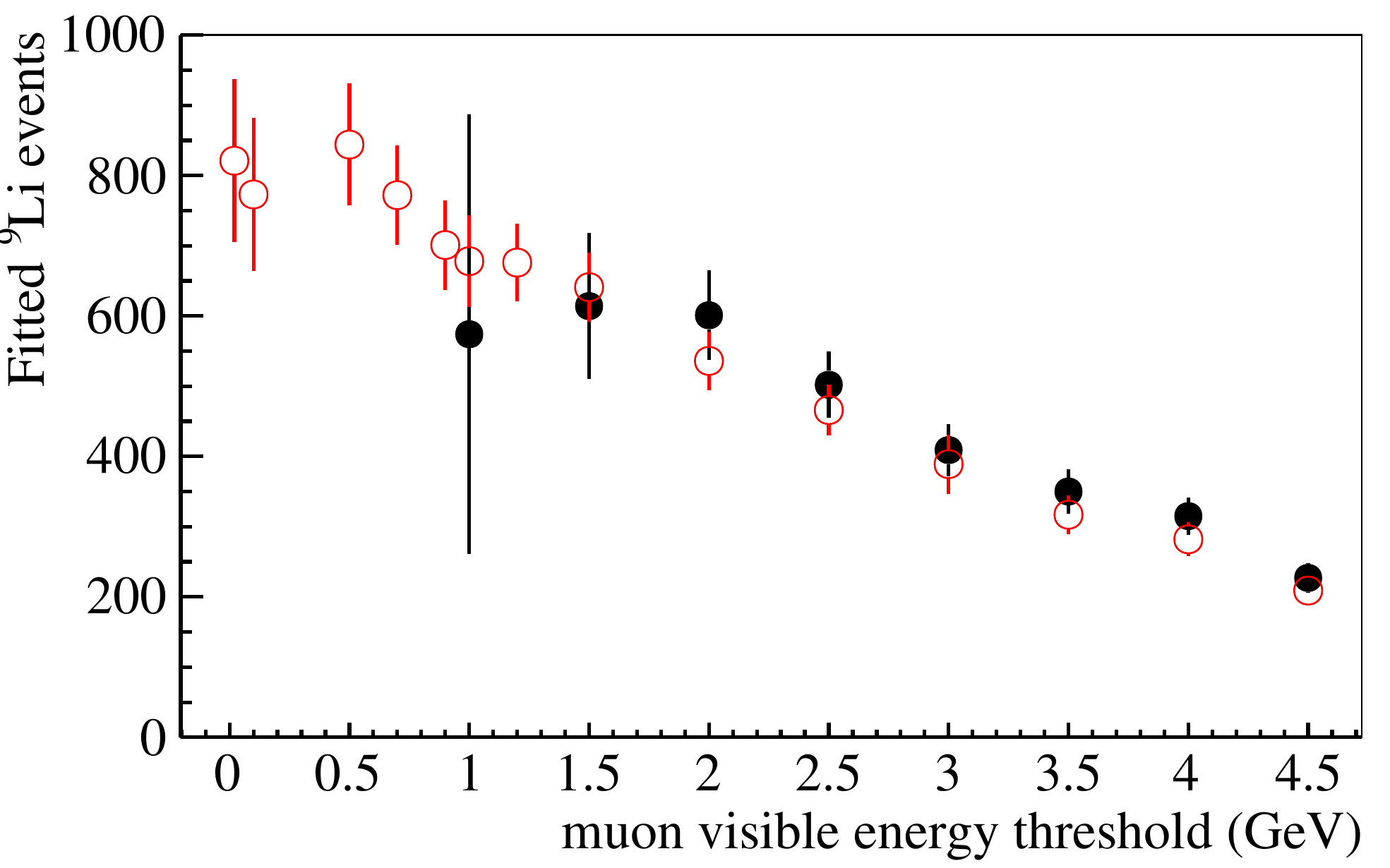}
\end{minipage}
\begin{minipage}[t]{0.5\textwidth}
\includegraphics[width=\textwidth]{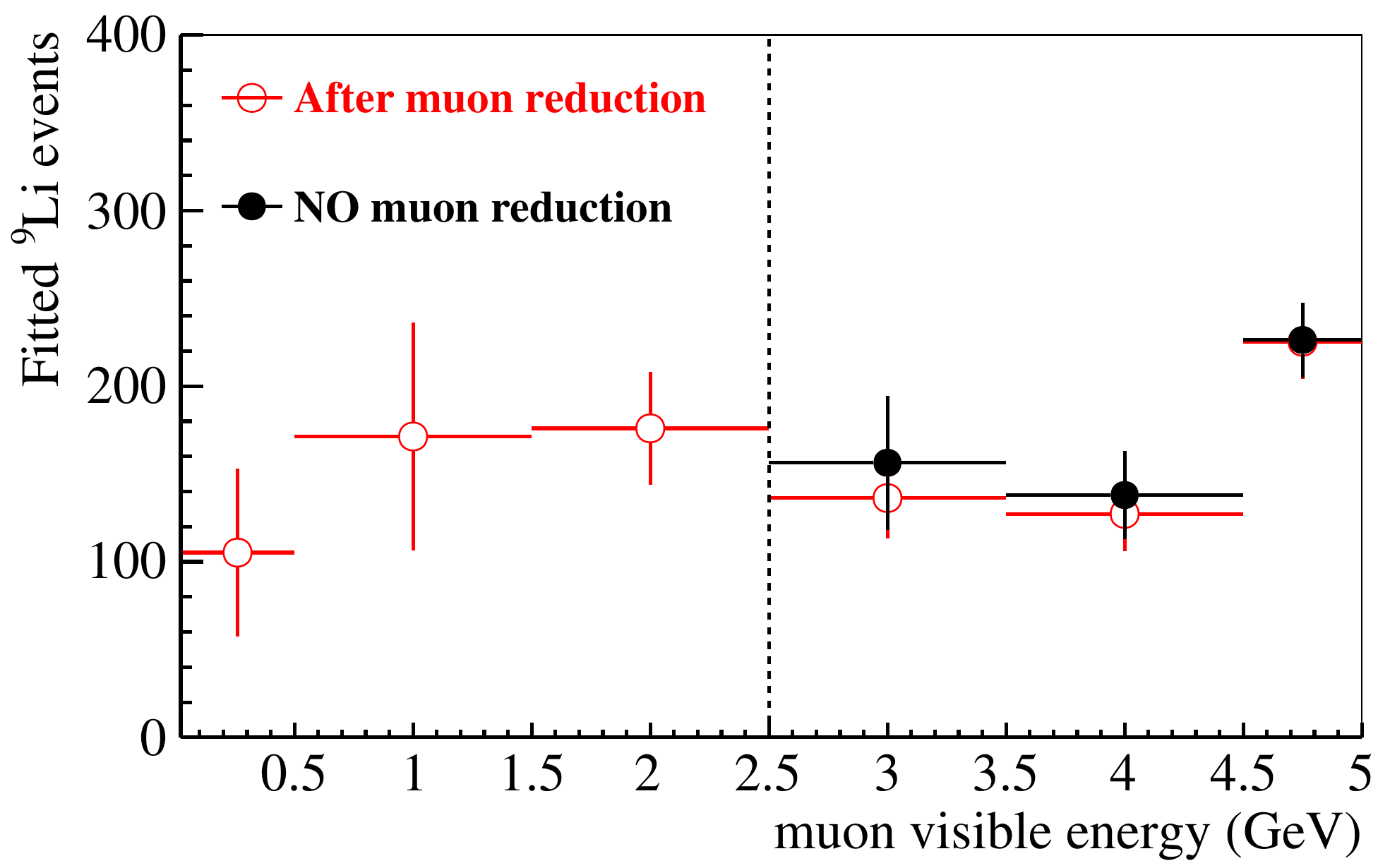}
\end{minipage}
\caption{\label{fig:dybLi} The fitted $^9$Li backgrounds versus the threshold (left) or different ranges (right) of the visible energy of the muons events passing the antineutrino detectors in Daya Bay EH1, with (red, open circle) and without (black, full circle) follow-on neutrons. }
\end{figure}

The background rate can be extracted from the distribution of time between each \nuebar candidate and the closest preceding muon~\cite{Wen:2006hx}. This method was found to have large uncertainty at high muon rate, because the time constants for the $\beta$-n emitters and \nuebar signal are nearly degenerate. Since most \Linine/\Heeight are produced by showering muons, reducing the minimum ionizing muons in the muon sample would enhance the time correlation between muon and \Linine/\Heeight. Daya Bay observed significant $\beta$-n production from muons with no associated shower in LS~\cite{Dayabay,An:2013uza}, by comparing muon samples with and without follow-on neutrons, based on the hypothesis that the \Linine/\Heeight production was most-likely accompanied with neutron generation, as shown in Figure~\ref{fig:dybLi}. This indicated that, although the \Linine/\Heeight yield was expected to be much lower for a muon not showering inside the detector, it was compensated by the much higher rate of non-showering muons.

The above approach significantly reduced the largest background error for Daya Bay. In the most recent Daya Bay result~\cite{An:2016ses}, more than half of the \Linine/\Heeight production followed the small fraction of muons with associated showers. The estimates for muons with lower reconstructed energy, were inconclusive due to degeneracy of the $\beta$-n and \nuebar time constants in the distribution relative to those muons. Instead, $\beta$-n production by these muons was estimated using the neutron-tagged muon sample. The neutron-tagging efficiency for $\beta$-n production was assumed to be the same for muons that identified as showering or non-showering. A systematic uncertainty of $\sim\,$40\% was assigned, according to the variation in the observed tagging efficiency versus the muon deposit energy with this assumption.

The periods of reduced power of the reactors, particularly the reactor-OFF period, provide good opportunities for directly measuring the background rate. Double Chooz had 7.24 days of reactor-OFF data~\cite{Abe:2014bwa}, and the tension between the prediction and observation was around 2$\sigma$, leaving no room for unknown backgrounds.

Excellent muon tracking can improve the rejection of \Linine/\Heeight, since both KamLAND data~\cite{KamLAND-spall} and simulations by FLUKA/Geant4 showed approximately exponential distribution of the lateral distance from the $\beta$-n emitter to its parent muon track. JUNO will equip $\sim\,$36,000 3-inch PMTs (see Section~\ref{sec:juno}), providing large dynamic range of muon energy deposit measurement and excellent tracking for both minimum ionizing muons and the showering muons.

\subsection{Muon Veto}

Besides sufficient overburden, a high efficiency and redundant active veto system should be well designed to reject the cosmogenic backgrounds. The moun system at each site of Daya Bay consists of a water pool acting as water Cerenkov detector, and a plane of resistive plate chambers (RPC) on top of it. In addition, each pool is divided into two independent water Cerenkov detectors: inner water system (IWS) and outer water system (OWS), optically isolated with Tyvek. Double chooz has a LS veto detector and a plastic scintillator tracker on top. KamLAND and RENO uses a water Cerenkov detector as veto. While the veto of KamLAND has a not very high efficiency, that of the $\theta_{13}$ experiments perform very well, e.g. Daya Bay IWS has an efficiency of 99.98$\pm$0.01\%. Having very good muon tracking is more important for JUNO to reject muon-induced long-live isotope backgrounds. The design will be described in Section~\ref{sec:junoVeto}.

\subsection{Issues with Chimney}

Double Chooz and RENO have a chimney for deploying the calibration sources. An outer veto covers the top of the detector tank and the chimney region except for the chimney barrel. Stopping muons which entered through and stopped inside the chimney, if failing to be identified by the veto system, could produce correlated backgrounds. Because the stopping muons in the chimney have a different hit pattern than a point-like source in the target, the likelihood with the best-fit vertex was used to reject such backgrounds. In Daya Bay, the automated calibration units are integrated on the top of antineutrino detector and submerged in water, thus the background induced by stopping muons is negligible.

\par
For the JUNO detector, a chimney on top of the acrylic sphere is designed for filling LS and operating the calibration systems. Based on the lessons on the chimneys of KamLAND, Double Chooz and RENO, JUNO will equip a light blocker at the chimney bottom to prevent the light generated inside the chimney from entering the main LS volume. In addition, the inner surface of chimney will be black. A detailed MC study indicated that the stopping muon background would be negligible.

\subsection{Others}
\label{sec:others}

\par
Many neutrino experiments observed problematic instrumental noises, namely PMT {\it flashers}, which produce spontaneous light emission due to the discharge on dynodes or PMT bases. The flashing PMTs were usually turned off, resulting in loss of light and additional non-uniformity. The flasher events in Daya Bay produced specific charge patterns that were distinct from physical events, thanks to the cylindrical arrangement of the PMTs. Thus they were easily identified and efficiently removed with a simple flasher identification variable~\cite{An:2013uza}. For future experiments, to avoid spontaneous flashing, particular care should be taken for the fabrication and potting of the PMT base circuits.

\par
A special correlated background existed in Daya Bay, due to the $^{241}$Am-$^{13}$C neutron source in each bell jar of the automated calibration units (ACU). A neutron emitted from the low rate ($\sim\,$0.5 Hz) Am-C neutron source~\cite{Gu:2015inc} could generate a $\gamma$-ray via inelastic scattering in the stainless steel vessel, then subsequently be captured on Fe-Cr-Mn-Ni. If both $\gamma$ rays from the scattering and capture processes enter the scintillating region, it could mimic a \nuebar signal. To mitigate this background, the $^{241}$Am-$^{13}$C sources in the two off-axis ACUs were removed from each of the Daya Bay far-site detectors. RENO suffered an accident that a tiny amount of $^{252}$Cf leaked out from the calibration source and contaminated both its detectors. It was caused by that an O-ring of an acrylic container coming loose due to aging, then the Gd-LS smeared in and out of the container~\cite{RENO:2015ksa,Seo:2016uom}. Since on average 3.7 neutrons are emitted from a $^{252}$Cf decay, a stringent multiplicity cut was applied to reject the $^{252}$Cf contamination background. Future experiments should take the lessons.

%% file: section_5.tex
%!TEX root = reactor_AR_main.tex

\section{FUTURE REACTOR NEUTRINO EXPERIMENT (JUNO)}
\label{sec:futureExp}

Discussions about the experimental feasibility of JUNO proposal (known as Daya Bay II then) started in 2008. The needed event statistics and energy resolution in fact requires a significant technology advancement on PMTs, LS and the engineering of the very large antineutrino detector. It was a huge challenge to predict what can be improved and where is the limit. A conceptual design of JUNO was completed in 2009 and a rigorous R\&D effort was started then. Now JUNO is in the construction phase. In the following we will discuss the design and R\&D results of JUNO, together with the construction status and the future plan.

\subsection{Detector Design}
\label{sec:junoDet}

\par
\begin{figure}[!htp]
  \centering
  \includegraphics[width=0.7\textwidth]{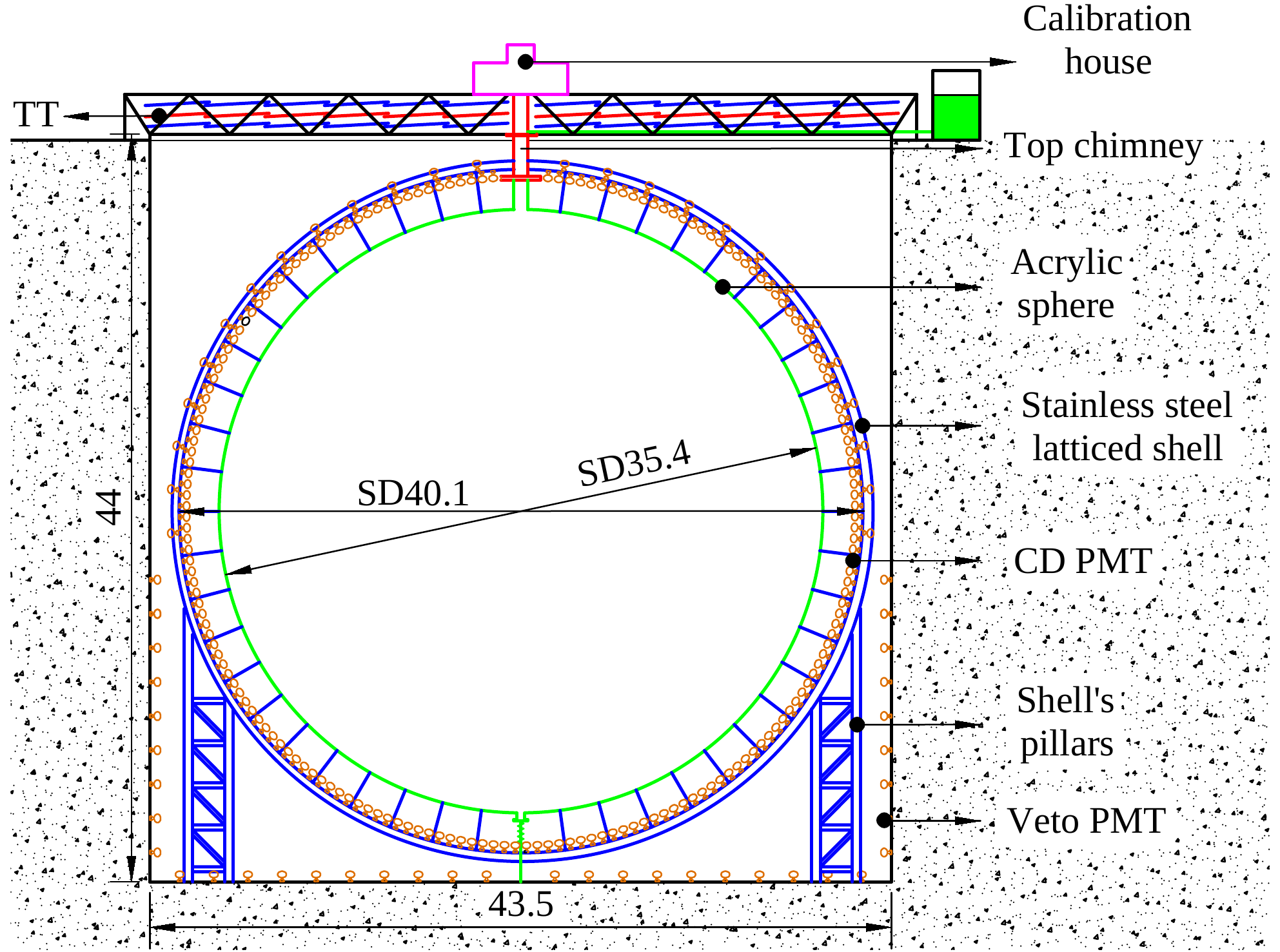}
  \caption{The layout of JUNO detectors.}
  \label{fig:junoDet}
\end{figure}

The JUNO detectors will be constructed underground with a vertical overburden of 700 m. JUNO has one antineutrino detector (namely the central detector), two redundant muon veto systems, complementary calibration systems and FADC readout electronics system. The detectors layout is shown in Figure~\ref{fig:junoDet}, and the details of key subsystems are discussed below.

\subsubsection{Central Detector}
\label{sec:junoCD}

\par
Enormous engineering challenges exist to build a 20 kton LS detector, which is 20 times larger than KamLAND. Four options had been discussed before 2014, then reduced to two, known as the ``Acrylic Sphere" option and the ``Balloon" option~\cite{Heng-ICHEP16}. The final choice was made in 2015 to be the ``Acrylic Sphere" option.

The final design of JUNO central detector consists of 20 kton low-background LS held by an acrylic sphere 35.4 m in diameter and 120 mm in thickness. The acrylic sphere will be fabricated on-site by bonding $\sim\,$260 pieces of large acrylic panels, and is supported by a latticed shell made of stainless steel. About 18,000 20-inch PMTs, instrumented on the latticed shell, look inward to the LS and provide a maximum photocathode coverage of 75\%. In the gaps between the 20-inch PMTs, around 36,000 3-inch PMTs will be installed, providing additional 3\% photocathode coverage, enhancing the capability and dynamic range of measuring cosmic muons, and serving as an independent calorimeter to calibrate the energy nonlinearity. The whole structure is submerged in a ultra-pure water pool (see Section~\ref{sec:junoVeto}).

For this ``Acrylic Sphere" design, the biggest challenge is to build and support the world's largest acrylic sphere. A reliable design has been achieved, after several years of prototyping and bonding tests, mechanical tests and FEA validation. Issues such as temperature variation, earthquake safety, single point failure analysis, installation process and PMT mounting, etc, have been extensively studied and detailed engineering are basically finished. The energy response has a complicated shape near the edge of acrylic sphere due to the internal reflection, thus a comprehensive calibration system is designed, as discussed in Section~\ref{sec:junoCal}. Good progress has been made on implosion protection of PMTs, highly reliable PMT potting, under-water electronics, and low Radon concentration in water.

In the abandoned ``Balloon" option, a transparent PA-film balloon holds the LS target, and is immersed in a LAB buffer which is contained in a stainless steel vessel (SSV). The leaks of such big balloon, most likely unavoidable, could be a show stopper. In the worst scenario, if the buffer LAB becomes scintillator, the radiation from PMTs and SSV will provide vast number of light noise overlapping with the \nuebar signal. The MC simulation indicated significant degradation in the energy resolution and decreased the NMO sensitivity by $\Delta\chi^2\sim4.7$. The design for chimney and calibration system is more complicated. The control of radioactivity contamination on such large surface ($\sim\,$4000 m$^2$) is non-trivial. Two 12-m-diameter balloons were prototyped to evaluate the design and fabrication, the details about the ``Balloon" option will be published soon.

\subsubsection{Calibration System}
\label{sec:junoCal}

\par
The absolute energy scale uncertainty is required to be \textless1\% for JUNO, thus a sophisticated calibration program is important. Based on the calibration experiences from Daya Bay~\cite{Gu:2015inc,Liu:2013ava,Liu:2015cra,Huang:2013uxa}, complementary calibration systems are being designed for JUNO: the automated calibration unit (ACU), cable loop system (CLS), guide tube control system (GTCS), and remotely operated under-liquid-scintillator vehicles (ROV) system. The ACUs are used to regularly deploy sources along the central axis, and similar technologies worked excellently at Daya Bay. The CLS system can deploy sources at different locations in one given vertical plane. The GTCS allows deploying sources along a given longitude of the acrylic sphere to calibrate the energy response at the LS edge. The ROV can scan the whole LS volume to do a 4$\pi$ calibration. Various gamma sources ($^{60}$Co, $^{40}$K, $^{54}$Mn, $^{137}$Cs), positron sources ($^{68}$Ge, $^{22}$Na) and neutron sources ($^{241}$Am-Be, $^{241}$Am-$^{13}$C, or $^{241}$Pu-$^{13}$C, $^{252}$Cf) are being investigated, and which system carries what sources is under evaluation.

\subsubsection{Muon Veto System}
\label{sec:junoVeto}

\par
The design of JUNO muon veto system has good redundancy and inherits many technologies from Daya Bay~\cite{Dayabay:2014vka}. A large water Cerenkov detector, 44 m in height and 43.5 m in diameter, surrounds the central detector. The stainless steel truss that supports the acrylic sphere automatically divides the water pool into two optically isolated zones: the inner water shield and the outer water Cerenkov detector with $\sim\,$2000 PMTs. The inner shield attenuates the radiation from PMTs and steel latticed shell, and the outer detector actively tags the muons. The minimum water thickness is $\sim\,$3.9 m to ensure sufficiently low fast neutron backgrounds. The target tracker of the OPERA experiment is re-used to cover the top of the water pool in a three layer arrangement. This external veto detector will provide optimum muon tracking and cross check with the central detector to ensure a highly efficient muon veto.

The performance of the 20-inch PMT is significantly affected by the Earth magnetic field, thus compensation coils are designed inside the water pool, to suppress the field by at least one order of magnitude.

\subsubsection{FADC Readout and Trigger}
\par
JUNO will record the full waveform for each PMT channel by using high resolution Flash ADC (FADC) with 1 GHz sampling frequency. Then the number of photoelectrons can be reconstructed by a de-convolution method based on fast fourier transform (FFT) technique, which resolves the hit time of each photoelectron from the measured waveform. This can largely remove the non-linear response from the electronics side. Given the vast number of PMTs and long distance from PMTs to the electronics room, a highly-integrated readout electronics is designed to be integrated in the PMT housing, to reduce the number of cables and noise. The PMT base circuit, analog frontend and digitizer are enclosed in the waterproof potting shell. In addition, a customized high voltage (HV) module is also integrated to provide the PMT bias voltage.

The capability of filtering low energy ($<0.2$ MeV) noise event, predominantly by the coincidence of PMT dark noise in the trigger time window, is important for studying supernova neutrinos and solar neutrinos. The multiplicity trigger, which utilizes the received number of over-threshold channels, can effectively and robustly reject the PMT dark noises. Furthermore, JUNO trigger system employed an online correction to the time-of-flight (TOF) of photons, with a TOF look-up table that is constructed by dividing the LS volume into voxels. Then the PMT hits are aligned into a 50 ns narrow window for triggering, allowing significant removal of noise and efficient triggering for physics events below 0.2 MeV. In the case of a supernova burst at 10 kpc from the Earth, it will register several thousands of events within $\sim\,$10 s in the JUNO detector. The readout system can work in pipeline mode to record all the waveforms.

\subsection{Physics Requirements and Critical Technology R\&Ds Toward the Design}
\par
Maximizing the collected photons is the most critical factor driving the JUNO R\&D programs and detector design. From the PMT side, the focus is on improving the quantum efficiency and collection efficiency of the 20-inch PMT and reducing the radioactivity of PMT glass. From the LS side, the primary focus is the optimal procedures for on-site purification to obtain more transparent and ultra-low radioactivity scintillator. Precise measurement of the Rayleigh scattering length and complete understanding of the light propagation in LS are of great importance in such a very large detector. In addition, many engineering challenges need be resolved to build the giant detector.

\subsubsection{Maximum Photocathode Coverage with High Efficiency 20-inch PMT}
\label{sec:mcp-pmt}
\par
It is known that large area PMTs have the best performance-to-price ratio, as in the case of SuperK. The best choice at that time was Hamamatsu R3600 20-inch PMT with an average quantum efficiency (QE) of $\sim\,$22\%, and collection efficiency (CE) of 70\%, resulting in a total photon detection efficiency of $\sim\,$15.4\%. This is in fact a factor of 2 short for JUNO. A completely novel design using a Microchannel Plate (MCP) in place of a dynode to amplify photoelectrons was proposed~\cite{Wang:2012rt}. This design has the transmission photocathode coated on the front hemisphere and the reflection photocathode coated on the rear hemisphere to form nearly 4$\pi$ viewing angle to enhance the photoelectron detection efficiency. The MCP has intrinsically high CE thanks to the simple assembly of the MCP structure.

Successful prototypes of the 20-inch MCP-PMT were made following such an approach in the past years. The photoelectron CE depends on the MCP open area fraction, the angular and energy distributions of electrons emitted from the photocathode, the potential difference between the PMT photocathode and the MCP surface, as well as the secondary electron emission from the MCP electrode. The structure and operational parameters of the 20-inch MCP-PMT have been optimized to improve the photoelectron CE.

The current 20-inch MCP-PMT achieved a total QE of $\sim\,$30\% at 400-420 nm, attributed as 26\% from the transmission photocathode and 4\% from the reflection photocathode, and a CE of $\sim\,$100\%~\cite{Qian-NNN16}. Given the total $\sim\,$180 tons of the PMT glass in the JUNO detector, low radioactivity is desired for the PMT glass bulb to reduce the accidental backgrounds. With extensive tests, the radioactivity of the MCP-PMT glass has been controlled to $\lesssim$1.2 Bq/kg $^{238}$U, \textless0.4 Bq/kg $^{232}$Th and \textless0.4 Bq/kg $^{40}$K, and expected to be further reduced. The detailed procedures for the radioactivity control will be published elsewhere. The other performances showed \textgreater3 peak-to-valley ratio, $\sim\,$30 kHz dark rate and $\sim\,$12 ns transit time spread (TTS, defined as FWHM of the transit time distribution). The relative large TTS is an outcome of maximizing the CE during the optimization of the MCP-PMT design.

The MCP-PMT was developed by IHEP in collaboration with NNVT (North Night Vision Technology). During this process, Hamamatsu improved the QE of their PMT to $\sim\,$30\% by adopting the so-called SBA technology, and the CE to $\sim\,$90\% using Box and Line dynode design. Therefore the two vendors have very similar products which greatly helped the JUNO procurement. Based on the performance and cost of the PMTs, and considered the risk mitigation, JUNO decided to purchase 3/4 of the 20-inch PMTs from NNVT and 1/4 from Hamamatsu. A manuscript that describes the procurement approach in detail will be published elsewhere.

\subsubsection{Transparent and Radio-pure Liquid Scintillator}
\label{sec:junoLS}

The JUNO LS will consist of linear alkyl benzene (LAB) as solvent, PPO as primary fluor and bis-MSB as wavelength shifter. There are two baseline requirements to the JUNO LS:
\begin{itemize}
  \item Optical transparency: $>$20 m attenuation length for 430 nm optical photons
  \item Radio-purity: $^{238}$U\textless10$^{-15}$ g/g, $^{232}$Th$<$10$^{-15}$ g/g, $^{40}$K$<$10$^{-16}$ g/g, $^{210}$Pb$<$10$^{-22}$ g/g
\end{itemize}
On-site purification is necessary to achieve the requirements on optical transparency and radio-purity. Based on the experience from Borexino~\cite{Benziger:2007aq}, the R\&D programs have been taken on four purification techniques and corresponding facilities: Al$_2$O$_3$ absorption columns, distillation, steam stripping, and water extraction.

The distillation process and Al$_2$O$_3$ absorption columns are effective for optical purification. Distillation is an equilibrium-staged process. It removes impurities based on compositional changes associated with phase changes, utilizing the differences in the equilibrium composition between liquid and vapor. The distillation process removes impurities that are less volatile than LAB, thus effectively improves the optical transparency of LAB. But it does not remove noble gas impurities. The Al$_2$O$_3$ absorption columns are most effective in improving the transparency of LAB. The attenuation length of purified LAB and LS has achieved 24 m and 20 m, respectively.

The processes of steam stripping and water extraction are effective in removing the radioactive impurities. Steam stripping is also an equilibrium-staged process, but it removes the impurities that are more volatile than LAB: Ar, Kr and Rn. Water extraction is based on the differences in equilibrium concentration between the organic liquid and water, and highly effective in removing radioactive metal impurities: U, Th, K, and Pb.

A LS pilot plant has been built in Daya Bay LS hall~\cite{Zhou-Neutrino16}. A new batch of LS will be produced to replace the LS of one Daya Bay antineutrino detector, then purified via online circulation. The source calibration data and physics data will be taken to evaluate the optical property and radioactivity of the non-purified and purified LS. Based on the performances of above purification approaches, the final LS purification scheme for JUNO will be determined.

During the R\&D of LS purification, the understanding to the Rayleigh scattering in LS significantly improved. The attenuation length ($\lambda_{att}$) of LS samples, typically measured by a 1-m tube apparatus with monochromatic light beam (430 nm), actually attributes to the real absorption ($\lambda_{abs}$) and the Rayleigh scattering ($\lambda_{Rayleigh}$), which can be described as $\lambda_{att}^{-1}=\lambda_{abs}^{-1}+\lambda_{Rayleigh}^{-1}$. The intrinsic limit for the measured LS attenuation length is set by the Rayleigh scattering. The precise measurements to the Rayleigh scattering length and the apparent attenuation length are very important to predict the photoelectron (P.E.) yield at JUNO. A recent work determined the Rayleigh scattering length of LAB at 430 nm to be $\sim\,$27 m~\cite{Zhou:2015gwa}, based on the Einstein-Smoluchowski-Cabannes formula that considers the non-zero depolarization ratio of the organic liquids. Given the achieved 20 m attenuation length, the real absorption length is estimated to be 77 m at 430 nm. These parameters are input to the JUNO MC.

\begin{figure}[!htp]
  \centering
  \includegraphics[width=0.6\textwidth]{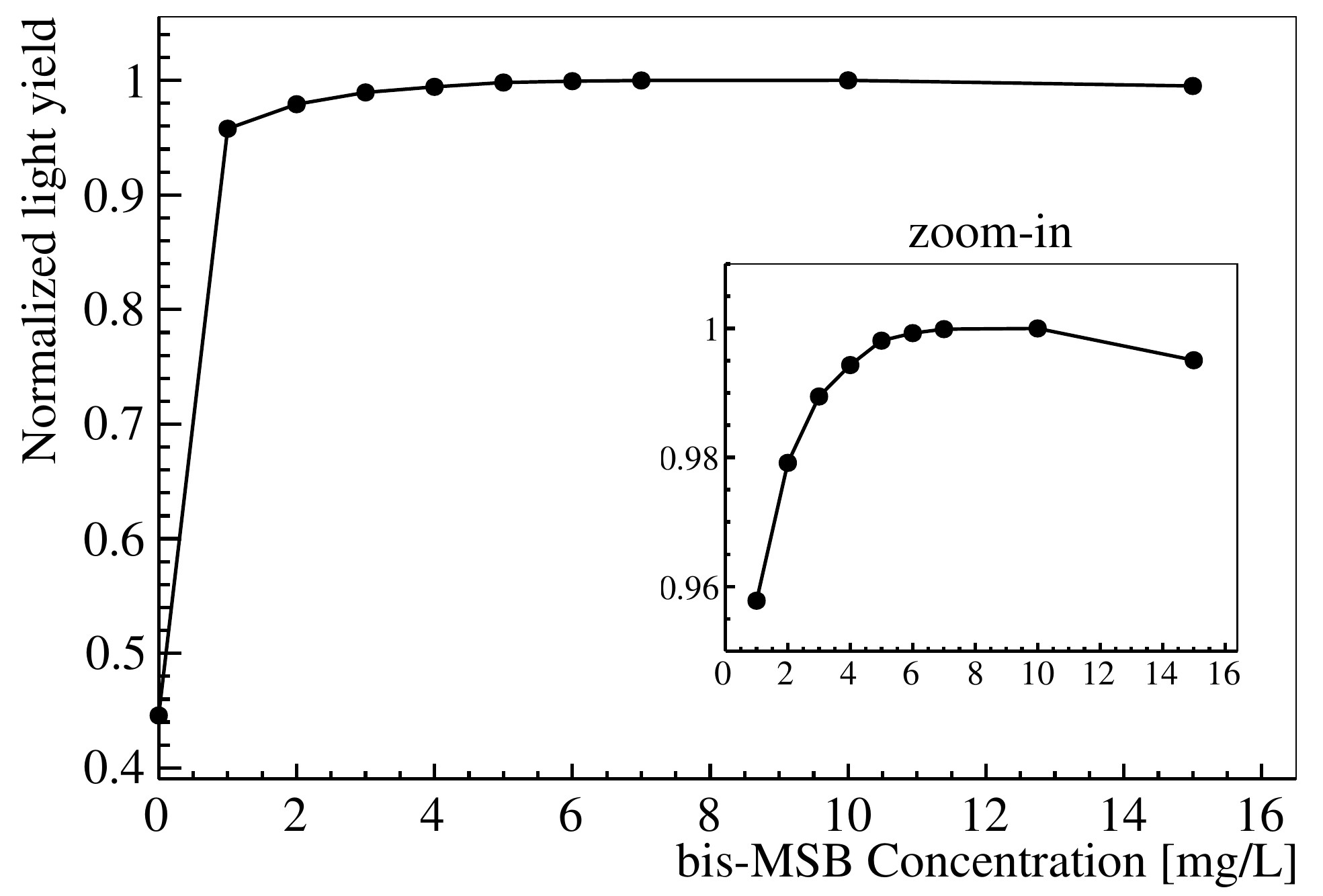}
  \caption{The dependence of P.E. yield on the bis-MSB concentration for a large LS detector like JUNO, which is predicted by the newly developed optical model with a fixed concentration of PPO.}
  \label{fig:lsopt}
\end{figure}

The JUNO LS is a ternary system, and the photo-absorption among each LS composition is a competitive process. The photon absorber, particularly the fluor molecule, has a certain probability to re-emit a new photon. Due to such complicated absorption and re-emission processes, the optimization of the LS recipe for large LS detectors cannot rely on laboratory measurements. Based on the extensive measurements~\cite{Feng:2015tka}: the molar attenuation coefficient of each LS composition and the quantum yield of each fluor, a generic optical model in Monte Carlo has been developed. Such model is capable of optimizing the LS recipe of JUNO, and Figure~\ref{fig:lsopt} is a preliminary result for optimizing bis-MSB concentration while the PPO concentration is fixed. With the LS pilot plant mentioned above, new LS with different concentrations of PPO and bis-MSB will be produced and filled into the same Daya Bay detector. With the well-understood detector and new calibration data, the optical model can be verified and further improved. A manuscript that describes this optical model will be published.

\subsubsection{Energy Resolution}
\label{sec:junoReso}
\par
The optical processes and the absolute light yield normalization for the JUNO MC was derived from the Daya Bay MC which agreed very well with data. Assuming JUNO has the same photocathode coverage as KamLAND (34\%), the same 20-inch PMT as Super-K (see Section~\ref{sec:mcp-pmt}), and the same LS optical properties\footnote{The Daya Bay LS also used LAB as solvent, thus the same $\lambda_{Rayleigh}=27$ m is assumed in the Daya Bay MC. The $\lambda_{att}$ of the Daya Bay LS is measured to be 15 m, resulting in $\lambda_{abs}=34$ m.} as Daya Bay, the simulated P.E. yield is only $\sim\,$110/MeV.

From the light collection point of view, the design and R\&Ds achievements from the above technical advancements are highlighted below
\begin{itemize}
  \item 20 m attenuation length for 430 nm optical photons
  \item 75\% photocathode coverage
  \item high photo-detection efficiency PMTs (\textgreater27\% at 420 nm)
\end{itemize}
With these new parameters as inputs, the simulation shows that the JUNO detector has a P.E. yield of $\sim\,$1250/MeV, corresponding to a statistical resolution of 2.8\%/$\sqrt{E}$. {Here we have assumed that the Daya Bay 8-inch PMT has the same collection efficiency as the new Hamamatsu 20-inch PMT.} The improvement on the P.E. yield breaks down to a factor of $\sim\,$2.7, $\sim\,$2.2, and $\sim\,$1.9 increase due to the improvements on the LS transparency, photocathode coverage, and PMT detection efficiency, respectively.

\par
The energy resolution can be parameterized as $\frac{\sigma}{E}=\sqrt{\frac{A^2}{E}+B^2+\frac{C^2}{E^2}}$, where $A$ is the stochastic term, $B$ is the constant term and $C$ is the noise term. The P.E. yield determines the stochastic term. The possible contributions to the constant term include residual non-uniformity, PMT charge response and electronics effects, and instability of detector response. The noise term is predominantly contributed by PMT dark noise. When taken these effects into account with typical parameters from Daya Bay, the projected energy resolution of the JUNO detector is 3.0\%/$\sqrt{E}$.

\subsection{Future possibilities}

Whether neutrinos are of a Majorana nature is of fundamental importance for particle physics. Currently the neutrinoless double-beta (\zeronubb) decay process is the only experimentally feasible and most sensitive probe to this question. If the \zeronubb decay happens, there must exist an effective Majorana neutrino mass term~\cite{Schechter:1981bd,Duerr:2011yh,Liu:2016oph}. Precise determination of neutrino oscillation parameters at JUNO can help to reduce the range of the \zeronubb decay half-life predictions by a factor of 2~\cite{Ge:2015bfa}. If neutrinos have an inverted mass ordering, \mbb will be greater than $\sim\,$0.015 eV  based on current and projected knowledge of neutrino mixing parameters.

Numerous efforts have been made in the past decades for searching for \zeronubb processes, yet no signal was observed. The current generation \zeronubb experiments ~\cite{Agostini:2016,Albert:2014awa,KamLAND-Zen:2016pfg} set an upper limit on the effective Majorana mass of $m_{\beta\beta}<$(61-165) meV (90\% C.L.). JUNO, as a ultra-low background LS detector with excellent energy resolution, provides good opportunity for searching for \zeronubb. The concept is to build a clean balloon to hold the \zeronubb target and insert into the central region of JUNO LS. Ultra-low background is the key for a \zeronubb decay search, thus dissolving $^{136}$Xe-enriched, purified xenon gas into LS is a good choice. Of course doping other elements such as $^{130}$Te is also possible as LS technology develops. With an in-depth study about the backgrounds, JUNO was found to have potential to reach a sensitivity of $T^{0\nu\beta\beta}_{1/2}>1.8\times10^{28}$ yr at 90\% C.L. with $\sim\,$50 tons of fiducial $^{136}$Xe and 5 years exposure~\cite{Zhao:2016brs}, where the corresponding sensitivity on the effective neutrino mass could reach \textless 5-12 meV, covering completely the allowed region of inverted neutrino mass ordering. This would be a major step forward with respect to the experiments currently under planning. Such \zeronubb searches are a possible future program after NMO determination.

%% file: section_6.tex
%!TEX root = reactor_main.tex
\section{SUMMARY}
\label{sec:conclusion}

Reactor neutrino experiments have played a great role in establishing the three-flavor neutrino oscillations. We have reviewed the past, current and future reactor neutrino experiments, explaining how we ended up JUNO from CHOOZ, Palo Verde, KamLAND to Daya Bay, Double Chooz, RENO, etc. The detector energy nonlinearity and the reactor neutrino spectrum, relative to the Daya Bay measurement, could possibly be determined to $\sim\,$1\% uncertainty. With the technical advancement on high detection efficiency 20-in PMT, long attenuation length liquid scintillator, and optimal detector design, the challenging 3\%$/\sqrt{E}$ energy resolution required by the neutrino mass ordering determination should be achievable. Reactor experiment will be a key player in determining neutrino mass ordering and precision measurement of $\theta_{12}$, $\Delta m^2_{32}$, and $\Delta m^2_{31}$ to sub-percent level.